\documentclass[journal]{IEEEtran}

\ifCLASSINFOpdf
   \usepackage[pdftex]{graphicx}
\else
   \usepackage[dvips]{graphicx}
\fi
\usepackage{cite}
\usepackage{filecontents}
\usepackage[cmex10]{amsmath}
\usepackage{array}
\usepackage[font=footnotesize]{subfig}
\usepackage{stfloats}
\usepackage{url}
\usepackage[normalem]{ulem}
\usepackage{psfrag}
\usepackage{dsfont}
\usepackage{amssymb}
\usepackage{ textcomp }
\usepackage{algpseudocode}
\usepackage{algorithm}
\usepackage{mathtools, cuted}
\usepackage[cmex10]{amsmath}
\usepackage{amsfonts}
\usepackage{xcolor}
\usepackage{amsmath,epsfig}
\usepackage{amsmath,epsfig,amssymb,mathrsfs,psfrag,epsf,enumerate,bm,color}
\usepackage{booktabs}
\usepackage{color}
\usepackage{tikz}
\usepackage{graphicx}
\usepackage{tikz}
\usepackage{lipsum}
\usetikzlibrary{shapes,arrows,automata,positioning}
\usepackage{anyfontsize}
\usepackage{comment}

\usepackage{hyperref}
\usepackage{soul}
\usepackage{stfloats}

\newcommand*\circled[1]{\tikz[baseline=(char.base)]{
            \node[shape=circle,draw,inner sep=2pt] (char) {#1};}}

\newcommand*\squared[1]{\tikz[baseline=(char.base)]{
            \node[shape=rectangle,draw,outer sep=2pt] (char) {#1};}}

\newtheorem{theorem}{Theorem}
\newtheorem{lemma}{Lemma}
\newtheorem{definition}{Definition}

\newtheorem{corollary}{Corollary}
\newtheorem{remark}{Remark}

\newcommand{\argmin}{\operatornamewithlimits{argmin}}

\newcommand{\overbar}[1]{\mkern 1.5mu\overline{\mkern-1.5mu#1\mkern-1.5mu}\mkern 1.5mu}

\usepackage{pgfplots}
\pgfplotsset{compat=newest}
\pgfplotsset{plot coordinates/math parser=false}
\newlength\fheight
\newlength\fwidth

\makeindex

\begin{document}

\title{ Energy and Social Cost Minimization for Data Dissemination in Wireless Networks: Centralized and Decentralized Approaches }

\author{
	\IEEEauthorblockN{
		Mahdi Mousavi  
		and 
		Anja Klein,~\IEEEmembership{Member,~IEEE}   
     	}
     	\thanks{Copyright (c) 2015 IEEE. Personal use of this material is permitted. However, permission to use this material for any other purposes must be obtained from the IEEE by sending a request to pubs-permissions@ieee.org.}
 	    \thanks{The authors are with Communications Engineering Lab, Technische Universit\"at Darmstadt, Germany. 	E-mail: \{m.mousavi, a.klein\}@nt.tu-darmstadt.de.}
 	    \thanks{Part of this paper has been published in \cite{Mousavi_ICC16}.}
} 
\maketitle

\begin{abstract}
We study multi-hop data-dissemination in a wireless network from one source to multiple nodes where some of the nodes of the network act as re-transmitting nodes and help the source in data dissemination.
In this network, we study two scenarios; i) the transmitting nodes \textit{do not} need an incentive for transmission and ii) they \textit{do} need an incentive and are paid by their corresponding receiving nodes by virtual tokens.
We investigate two problems; P1) network power minimization for the first scenario and P2) social cost minimization for the second scenario, defined as the total cost paid by the nodes of the network for receiving data.
In this paper, to address P1 and P2, we propose centralized and decentralized approaches that determine which of the nodes of the network should act as transmitting nodes, find their transmit powers  and their corresponding receiving nodes.
For the sake of energy efficiency, in our model, we employ maximal-ratio combining (MRC) at the receivers so that a receiver can be served by multiple transmitters.
The proposed decentralized approach is based on a non-cooperative cost-sharing game (CSG).
In our proposed game, every receiving node chooses its respective transmitting nodes and consequently, a cost is assigned to it according to the power imposed on its chosen transmitting nodes. 
We discuss how the network is formed in a decentralized way, find the action of the nodes in the game and show that, despite being decentralized, the proposed game converges to a stable solution.
To find the centralized global optimum, which is a benchmark to our decentralized approach, we use a mixed-integer-liner-program (MILP).
Simulation results show that our proposed decentralized approach outperforms the conventional algorithms in terms of energy efficiency and social cost while it can address the need for an incentive for collaboration.
\end{abstract}

\IEEEpeerreviewmaketitle

\section{Introduction} \label{sec:Intro}
 
With the growing integration of communication networks, the Internet of things (IoT) and sensor networks under the umbrella of 5G, there is a  need for increasing the capacity of future dense wireless networks \cite{EricssonReport_18}.
Multi-hop communications is an efficient technique that can increase the capacity of wireless networks and improve the reliability of communications.

This paper specifically studies multi-hop broadcast where one source disseminates its data to multiple receivers.
The data here could be a file, a software update, a video stream, etc. 
In this multi-hop network, in order to decrease the transmit power required at the source for serving all the receiving nodes or to increase the coverage area, some of the nodes may need to re-transmit the source's data.
We investigate two scenarios in this network.
In the \textit{first scenario}, we assume that the re-transmitting nodes are willing to collaborate with the source in data dissemination.
Such a scenario may apply to IoT and sensor networks.
In the first scenario, the objective is to minimize the power required in the network for data dissemination.
In the \textit{second scenario}, however, we assume that the transmitting nodes require an incentive from their corresponding receiving nodes in exchange for transmitting the data to them.
Such an assumption is vital for  situations  in which human users play an active role in data dissemination.
Hence, in the second scenario,  we aim at minimizing the social cost, defined as the summation of the costs that the nodes pay in exchange for receiving the source's data \cite{MAS08}.
The payment in this scenario is via virtual tokens, assumed to be available at the nodes.

Due to the broadcast nature of the wireless medium, serving multiple nodes by multicast transmission is more energy-efficient than employing unicast transmission where every receiving node requires a separate transmission.
Multicast transmission can improve energy efficiency and the capacity of the future generation of wireless networks, for instance, in machine-to machine-transmissions  \cite{Condoluci_Access16} and video streaming \cite{Argyriou_Network17}.
Since there may exist multiple transmitting nodes in our network, a receiving node is able to receive multiple copies of the disseminated message.
The receiver  in this case, instead of relying on one transmitter, can combine the received copies in order to decode the message.
This helps in network energy-efficiency. 
Further, in a wireless device, two types of modules are involved in message transmission; passive and active modules.
The passive modules such as digital-to-analog converter, mixer, etc., require a fix power for proper operation \cite{Cui_TWC05}.
We refer to the power required for these modules as the circuitry power.
In contrast, the power required at the active module, that is, the amplifier, is not fixed and varies depending on the channel gain between a transmitter and a receiver.
This power is referred to as the radio link power.
Despite the fact that the circuitry power is not negligible compared to the radio link power \cite{Cui_TWC05}, it is often ignored for multi-hop communications and merely the radio link power is considered for optimization.
For the sake of energy-efficiency, in the present work, 
i) we consider multicast transmission so that multiple receiving nodes can be served by a common transmitter, 
ii) we let every receiving node  be served by multiple transmitting nodes and in this case,  maximal ratio combining (MRC) is used to combine the copies of the message  received at a receiver
 and 
iii) besides the radio link power, we take  into account the circuitry power for multi-hop broadcast and show that our model leads to form a network that requires a lower power for data dissemination compared to the circuitry power-agnostic approaches.

In order to design a decentralized mechanism, for both the first and the second scenarios, we use game theory.
A non-cooperative cost sharing game (CSG) is proposed in which every receiving node chooses one or more than one node among the nodes of the network as its respective transmitting node.
In addition,  the receiving node determines the transmit power of each of its chosen transmitting nodes such that the receiving node receives the message with a required minimum signal to noise ratio  (SNR).
As the consequence of the receiver's decision, a  cost is assigned to it.
By using a CSG, in a multicast transmission where a transmitting node has multiple receivers, the cost paid to the transmitter is shared among its receivers by a so-called cost sharing scheme.
The cost in the first scenario is an artificial cost, used as a tool by which a node merely finds the best transmitting nodes for itself in order to minimize the network power, whereas, in the second scenario, the cost of the node determines the cost, i.e.,  the number of tokens, that needs to be paid by the receiver to each of its chosen transmitters.

In both scenarios, the cost of a receiving node is defined as a function of the total transmit power of its chosen transmitting nodes, including the radio link and the circuitry powers.
Such a design is clear to be beneficial for the first scenario as we aim at network power minimization.
The reason for taking such an approach for the second scenario is the importance of battery-life for the end-users.
Studies show that energy consumption is one of the main concerns of the users when it comes to collaborative transmission \cite{shatri_access19}.
Hence, in the second scenario, the transmitting nodes are also paid by their respective receiving nodes based on the power they use for serving their receivers.
The fundamental difference between the CSG proposed for each of the scenarios is the employed cost sharing scheme.
In the first scenario, we use the marginal contribution (MC) cost sharing scheme as it is suitable for network power minimization \cite{Mousavi_TWC18}.
In the second scenario where every receiver pays a price in exchange for receiving the message, having a fair cost allocation is critical, especially when multiple receiving nodes are served by a common transmitting node via multicast.
In this scenario, we employ the Shapley value (SV) which is known as the fairest cost allocation scheme \cite{MAS08}.

Our proposed game is iterative where the nodes take their action one after another until the convergence of the game to a point at which none of the nodes can reduce its cost given the actions of the others, called the Nash equilibrium (NE) point.
The proposed game is shown to be a potential game for which the existence of at least one  NE  is guaranteed.

In addition to the decentralized approach, we find the centralized optimum network configuration by a mixed-integer linear program (MILP).
The optimum centralized solution serves as the benchmark for our proposed decentralized approach.
The centralized MILP formulation that we provide can be used for both scenarios by a minor change.

The rest of the paper is organized as follows. 
Section \ref{sec:StateArt} presents the related work  and our main contributions.
Section \ref{sec:SysModel} describes the network model and formulates the problem. 
The proposed decentralized algorithm and details of the game are  explained in Section \ref{sec:MRCgame}. 
Section \ref{sec:MRCMILP} provides the global optimum via an MILP.
Performance analysis is presented in Section \ref{sec:simul_multiPN} and finally, Section \ref{sec:summ_mPN} concludes the paper.

\section{State of the Art and Main Contributions }
\label{sec:StateArt}

Energy efficiency is one of the main challenges in multi-hop communications. 
Broadly speaking, the algorithms for multi-hop data dissemination are either centralized \cite{BIP, MY04} or decentralized \cite{Bader_ITJ16_iot, Alex15}.
One of the well-known centralized heuristics for minimum-power multi-hop broadcast is proposed by the authors in \cite{BIP}, called the broadcast incremental power (BIP) \cite{BIP}.
The BIP is a greedy algorithm, starts from the source and adds the nodes one by one to the network.
Being centralized is the main drawback of the BIP.
The dependency of the centralized approaches on a central entity makes them difficult to implement and vulnerable against the failure of the connection between the nodes and the central entity.
Hence, decentralized approaches are more suitable for such applications.
While numerous centralized algorithms have been proposed by researchers for minimum-power multi-hop broadcast, there does not exist much work on decentralized approaches.
Recently, decentralized solutions based on blind \cite{Bader_ITJ16_iot} and probabilistic \cite{Zeng_TVT18_vanet} approaches are proposed where in the former case, the nodes are unaware of the presence of each other, and in the latter one, the reception of the data is not guaranteed.
Since the transmissions by such approaches are unreliable, they cannot be used for streaming applications or payment-dependent transmissions.

The authors of  \cite{MY04} propose a greedy centralized heuristic that exploits MRC in multi-hop broadcast and we refer to it as GreedyMRC.
Like the BIP, it starts from the source and in every iteration, it finds the node which is the best transmitter in the network in terms of the energy required for transmission of the message to others.
Every transmitting node uses a separate time-slot for transmission and the nodes of the network receive the message over multiple time-slots  and combine them via MRC  until they accumulate the minimum SNR required for decoding the message.
Although GreedyMRC exploits MRC, it is centralized, does not address the incentive issue and ignores the circuitry power.
 
One of the main drawbacks of the existing algorithms for multi-hop broadcast is that they usually consider the nodes as passive and ready-to-collaborate entities.
In practice, and more specifically in the future generation of wireless networks, the users play an active role in data dissemination and require a proper incentive for collaboration \cite{Wang_TWC18_iot}. 
 In the past decade, many works have investigated various wireless communications and networking scenarios from a game-theoretic perspective, however, fewer works have studied incentive mechanisms.
Incentive mechanisms can be employed for sharing an already-cached data \cite{Wang_TWC18_iot} \cite{Mousavi_access19},  in store-carry-forward applications \cite{Cai_SCF_tvt_16}, base-station-assisted device-to-device communications \cite{Aditya_arxiv19}, distributed big data processing \cite{Zheng_TSPL17}, and network utility maximization \cite{Gao_TVT18}.
For instance, in \cite{Gao_TVT18}, the authors study network utility maximization in a network where a central entity has its own objective. They propose  mechanisms to motivate truthful reporting from the users to the central entity, about their personal preferences, so that  proper incentives can be  provided for the users to adopt the solution that the central entity prefers.
In \cite{Aditya_arxiv19}, an incentive mechanism is proposed within the framework of an auction  for selecting relay nodes in a cellular network who help the base station in transmitting data to other nodes.

The social cost, which we aim at minimizing in the second scenario, can be seen as a network utility.
In comparison to most of the existing works on network utility maximization, e.g., \cite{Aditya_arxiv19} and \cite{Gao_TVT18}, our proposed algorithm is fully decentralized.
It addresses the incentive design issue while at the same time it controls the network power.
Moreover, to minimize the social cost, we not only employ the fairest cost allocation scheme, that is, the SV, but also we show that the convergence of our designed decentralized algorithm with the SV is always guaranteed.
One of the open issues of the existing works is that they do not address how the incentives need to be calculated for each individual user.
A variety of rewarding schemes have been proposed by researchers, to be used as incentives, which rely either on a tit-for-tat strategy or reputation mechanism \cite{FlopCoin_TMC18}  or virtual currency \cite{Iosifidis_TON17}. 
With the growing popularity of cryptocurrencies in recent years, the latter strategy, that is, the use of virtual currency, is attracting significant attention \cite{FlopCoin_TMC18, localCoinJ}.
For instance, recently, the authors in \cite{FlopCoin_TMC18} proposed a virtual currency to be used for computation offloading in ad-hoc networks in which the users offload their tasks to other remote entities. 
A virtual currency similar to the one proposed in  \cite{FlopCoin_TMC18} can be employed in our network, however, in the present work, our focus is on the network formation rather than a token or a virtual currency design.

Due to the importance of user collaboration in the future generation of communication networks, it is vital to find a more comprehensive algorithm, capable of addressing both the energy efficiency and incentive issue for multi-hop broadcast.
This, to the best of our knowledge, has not been addressed in the literature.
Game theory, as a powerful mathematical tool, has been widely used for decentralized optimization of multi-user networks \cite{Li_IoTJ19, marden_welfare13, Mousavi_ISWCS15}. 
For instance, the authors of \cite{Li_IoTJ19} use  game theory in a content caching scenario where they aim to select important users and allocate content files to the storage of these selected users.
In \cite{Mousavi_ISWCS15}, we studied energy-minimization in multi-hop broadcast using a cost-sharing game.
In  \cite{Mousavi_PIMRC15}, we showed that using MRC at the receivers improves the performance of the network, however, the proposed algorithm for decision making at the nodes is a heuristic algorithm which does not necessarily find the optimum decision of the nodes.
We later showed in   \cite{Mousavi_ICC16} that the decision-making problem at the nodes can be solved optimally by a linear program in contrast to the heuristic approach of  \cite{Mousavi_PIMRC15}.
In comparison to \cite{Mousavi_ICC16}, in the present work, we improve our work in \cite{Mousavi_ICC16}, where a decentralized approach merely for social cost minimization was proposed.
Here, we propose a general framework for data dissemination which can be employed for both network power and social cost minimization.
We will show that in both cases, the cost function of the nodes can be modeled by a piece-wise linear function and the decision-making problem at a node can be solved via an MILP.
Moreover, in comparison to \cite{Mousavi_ICC16}, the power model that we consider for the nodes in the present work includes the circuitry power for both transmission and reception which makes the network power model much more realistic. 
We show that the circuitry power needed for signal reception, in particular, has a significant impact on the optimization problem.
We also, in the present work, discuss how the nodes can use a shared channel for transmission and exploit the MRC technique. 
Finally, to complete the work, we provide an MILP that finds the centralized global optimum configuration for both scenarios.

Briefly, the main contributions of	 this paper are as follows:

\begin{itemize}

 \item 
For  multi-hop broadcast we study  network power minimization when the transmitting nodes do not need an incentive for transmission and social cost minimization when having an incentive for the transmitting nodes is mandatory. When an incentive is required, we also address the fairness in cost allocation.
 
\item
Our algorithm supports multicast transmission and further, exploits the MRC technique by which the receiving nodes can be served by multiple nodes, instead of relying on one transmitting node for data reception.

\item
Besides the radio link power,  our model considers the circuitry power of the nodes for both transmission and reception of the message.
By such a model, the results obtained for the network power and the social cost are much more realistic compared to the existing works on multi-hop broadcast.

 \item
In comparison to \cite{Mousavi_ICC16} that provides a decentralized approach for social cost minimization in multi-hop broadcast, here, we propose both centralized and decentralized solutions  which can be employed for minimization of the network power as well as the social cost.

\item
We provide an MILP formulation that finds the global optimum configuration for both scenarios as a benchmark.

\end{itemize}

\section{Network Model and Assumptions} \label{sec:SysModel}
\subsection{Transmission and power models}
\begin{figure} [!t]
   \centering
   \includegraphics[width=.6   \columnwidth]{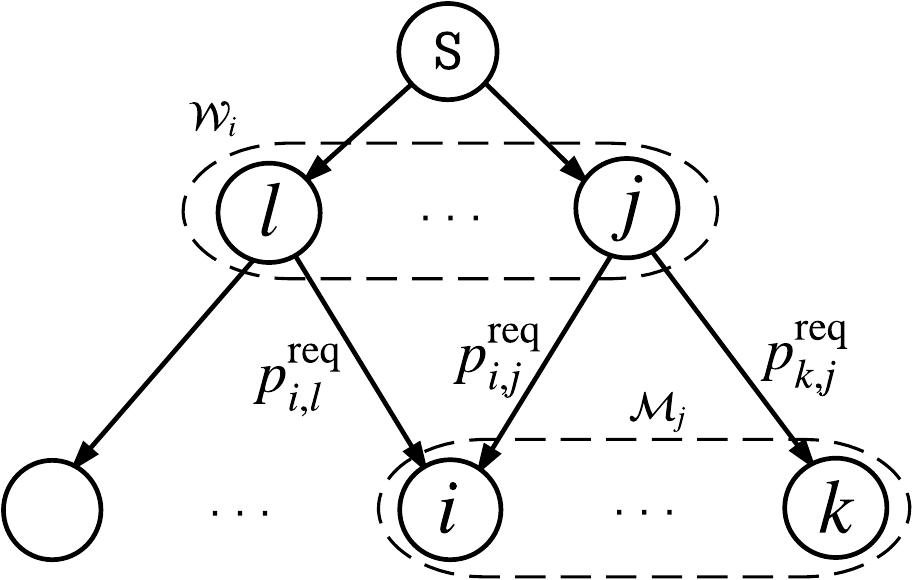}
   \caption{A sample network. In this network, node $i$ decides to be served by more than one PN, that is, nodes $l$ and $j$. }
   \label{fig:sysmodel_mrc}
\end{figure}

We consider a  network composed of $N+1$ wireless nodes with random locations in a two-dimensional plane; a source $\mathrm{S}$ and a set  $\mathcal{P} $ of $N$ receiving nodes.
The nodes in $\mathcal{P}$ are interested in receiving the source's message.
We denote the set of all the nodes of the network by $\mathcal{Q} = \mathcal{P} \cup \{\mathrm{S}\}$.
Every node $j \in \mathcal{Q}$ is equipped with an omnidirectional antenna and has an amplifier power constraint $p^\mathrm{max}_j$ for transmission over a radio link, and hence, its coverage area is limited.
For the sake of simplicity, we omit the efficiency coefficient of the amplifier of the transmitters.
In a transmission from  a transmitting node $j \in \mathcal{Q}$ to a receiving node $i \in \mathcal{P} \backslash \{j\}$, nodes $j$ and $i$ are called  the parent node (PN) and the child node (CN), respectively.
The PNs transmit either by unicast or multicast if they have one or more than one intended receivers, respectively.
The transmission flow from the source to other nodes of the network forms a directed acyclic graph (DAG), a graph in which there is at least one directed path from the source to all the nodes without any cycle, see Fig. \ref{fig:sysmodel_mrc}.

Every node $i \in \mathcal{P}$ chooses its PNs as well as the radio-link power of each of its chosen PNs. 
By  combining the  signals received from its chosen PNs, a CN can successfully decode the message if the accumulated SNR is at least equal to a threshold value $\gamma^\mathrm{th}$.
The radio link power used by PN $j$ to serve CN $i$, requested by the CN,  is denoted by $p_{i,j}^\mathrm{req}$. 
We define the \textit{action} of a node  as the set of tuples composed of the PNs that the node chooses along with the corresponding requested powers as
\begin{equation}
\label{eq:mrc:decision}
\bm{a}_i = \left\lbrace (j, p_{i,j }^\mathrm{req} ) | j \in \mathcal{A}_i, p_{i,j }^\mathrm{req} \in [0,p_j^\mathrm{max}] \right\rbrace ,
\end{equation}
in which $\mathcal{A}_i$ is the action set of the CN $i$ from which it chooses its PNs.
The set of actions of all the nodes is defined as 
\begin{equation}
\label{eq:mrc:actionset}
\bm{a} := \{\bm{a}_i | i \in \mathcal{P}\}.
\end{equation}
Given the transmit power of a transmitter, the SNR at a receiving node depends on the channel attenuation between the transmitter and the receiver as well as the noise power.
The SNR  at receiving node $i$ in a unicast transmission is defined as
\begin{equation}
\label{eq:snrdef}
 \gamma_{i,j} (p_{i,j}^\mathrm{req}) = \frac{  p_{i,j}^\mathrm{req} g_{i,j}  }{ \sigma^2 }.  
\end{equation} 
in which $g_{i,j}$ and $\sigma^2$  are the channel gain between nodes $i$ and $j$ and the noise power, respectively.
We assume that the transmitting nodes in the this network have their own channel for transmission and intra-network interference can be neglected \cite{Mousavi_TWC18}.
Further, we assume that the statistical properties of the channel remain unchanged during the data dissemination and the channel gains of the links between a given node and its neighboring nodes are known at the node.

The set $\mathcal{N}_i$ of the neighboring nodes of node $i \in \mathcal{P}$ is defined as the nodes which can provide the minimum SNR required for decoding the message at node $i$ via unicast considering their maximum radio-link power.
More precisely,
\begin{equation}
\label{eq:neighb}
\mathcal{N}_i = \{j | j \in \mathcal{Q}_i, \gamma_{i,j} (p_j^\mathrm{max}) \geq \gamma^\mathrm	{th}\} 
\end{equation}
in which $\gamma_{i,j} (p_j^\mathrm{max})$ is defined in \eqref{eq:snrdef}.
Although with MRC  a CN is able to accumulate the signals transmitted from the PNs which are not in its   neighboring area, we restrict the action set of a CN to the nodes in its neighboring area.
This is because  overhead information needs to be transmitted in practice between a PN and a CN in order to find the multicast receiving group, the cost of the CNs,  etc., and this implies that the PN must be in the neighboring area of the CN.
The action set of node $i$ in this section is defined as the set of neighboring nodes of node $i$ whose distance from the source, in terms of the number of hops, is not larger than that for node $i$.
More precisely, it is defined as
\begin{equation}
\label{eq:mrc:action_set}
\mathcal{A}_i = \{j | j \in \mathcal{N}_i, h_j \leq h_i, h_j \neq \infty\} 
\end{equation}
in which $ h_j$ is called the hop-rank of node $j \in \mathcal{Q}$, representing  the number of hops from the source to node $j$.
Denoting by $\mathcal{W}_i \subseteq \mathcal{A}_i$ the set of PNs of CN $i$, the hop-rank of  CN $i$   is obtained by 
\begin{equation}
\label{eq:hop_count_def}
h_i  = 
\begin{cases}
 \max_{j \in \mathcal{W}_i} \left\lbrace h_j \right\rbrace +1, &\text{if } \mathcal{W}_i \neq \varnothing \\ 
 \infty, &\text{if } \mathcal{W}_i = \varnothing,
\end{cases}
\quad \forall i \in \mathcal{P}, 
\end{equation}
which indicates that  the hop-rank of a node depends on the maximum of the hop-ranks of its PNs if $\mathcal{W}_i$ is not an empty set. 
\textit{Initially}, we set $h_\mathrm{S} =0$ and $h_j = \infty$ for all $j \in \mathcal{P}$.
Note that, $h_j \neq \infty$ in \eqref{eq:mrc:action_set} indicates that only a node that is already connected to the network and its hop-rank is not $\infty$ can be selected as a PN of a CN.
In fact, the first receiving node, say node  \circled{1}, can only choose the source node as its PN since the hop-rank of all the nodes except the source is $\infty$.
The next node, say node \circled{2}, now has two potential PNs; the source node with $h_\mathrm{S} = 0$ and node \circled{2} with $h_1 =1$.
This process continues until all the nodes join the network, however, they can later change their decision.
Based on the definition of the action of a node  in \eqref{eq:mrc:decision}, we observe that each action contains two sub-actions.
We call them the PN set and the power request set of node $i$ and define them as 
\begin{equation}
\label{eq:mrc:pn}
\mathcal{W}_i = \left\lbrace j \vert p_{i,j }^\mathrm{req} > 0, j \in \mathcal{A}_i \right\rbrace 
\end{equation}
and \vspace{-2mm}
\begin{equation}
\label{eq:mrc:pow_req}
\bm{p}_i^\mathrm{req} = \left\lbrace p_{i,j }^\mathrm{req} \vert  p_{i,j }^\mathrm{req} > 0, j \in \mathcal{A}_i \right\rbrace ,
\end{equation}
respectively.
The number of PNs chosen by CN $i$ is denoted by $W_i = |\mathcal{W}_i|$ and $W^\mathrm{max}$ is the maximum number of allowed PNs.
In Fig. \ref{fig:sysmodel_mrc}, $\mathcal{W}_i = \{l,j\}$.
From the perspective of a transmitting node $j \in \mathcal{Q}$, we define the set of its CNs as 
\begin{equation}
 \label{eq:mrc:CNset}
 \mathcal{M}_j = \left\{ i \big\arrowvert  p_{i,j}^\mathrm{req} > 0, \forall  i \in \mathcal{N}_j \right\}.
\end{equation}

We also denote the set of radio-link power requests \textit{received} by node $j$ from its neighboring nodes as 
$\bm{p}_j^\mathrm{rcv} = \left\lbrace p_{i,j }^\mathrm{req} \vert  i \in \mathcal{M}_j \right\rbrace  $
and
$\bm{p}_{-i,j}^\mathrm{rcv} := \bm{p}_j^\mathrm{rcv} \backslash \left\lbrace p_{i,j}^\mathrm{req} \right\rbrace $
represents $\bm{p}_j^\mathrm{rcv}$ without $p_{i,j}^\mathrm{req}$.
The \textit{radio-link power} of PN $j$ in a multicast transmission is given by
\begin{equation}
\label{eq:p_radio}
p_j^\mathrm{Tx}(\bm{p}_j^\mathrm{rcv}) = \underset{i \in \mathcal{M}_j}{\max}  \left\lbrace p_{i,j}^\mathrm{req} \right\rbrace .
\end{equation}

As mentioned in Sec. \ref{sec:Intro}, the hardware of a wireless transmitter consists of several passive modules which all require  a   power for proper operation.
We refer to this power as the \textit{circuitry power} of a node and denote the circuitry power of a node for transmission and reception by $p_j^\mathrm{ct}$ and $p_j^\mathrm{cr}$ for all  $j \in \mathcal{Q}$, respectively.
The circuitry power of a transmitter can be assumed as a fixed value.
Hence, the \textit{sum transmit power} of a PN $j$ is obtained by 
\begin{equation}
\label{eq:mrc:ptx_def}
P_j^\mathrm{Tx} (\bm{a}) := P_j^\mathrm{Tx} (\bm{p}_j^\mathrm{rcv}) = 
\mathds{1}_j(\mathcal{M}_j) \left(  p_j^\mathrm{ct} + \underset{i \in \mathcal{M}_j}{\max}  \left\lbrace p_{i,j}^\mathrm{req} \right\rbrace \right), 
\end{equation}
in which $\mathds{1}_j(\mathcal{M}_j)$ indicates if node $j$ acts as a transmitting node, i.e., $\mathds{1}_j(\mathcal{M}_j) = 1$ if $\mathcal{M}_j \neq \varnothing$ and $\mathds{1}_j(\mathcal{M}_j) = 0$, otherwise.
Using  MRC, the aggregated SNR experienced by CN $i$ is calculated by   
\begin{equation} 
\label{eq:MRC_Const}
\gamma_{i \vert_\mathrm{MRC}} ( \bm{p}_i^\mathrm{req} ) := \sum_{j\in\mathcal{W}_i} \frac{  p_{i,j}^\mathrm{req} g_{i,j}} { \sigma^2}. 
\end{equation}

The channel access scheme  employed in this network is composed of two sections.
The first section is a random-access channel (RACH) used by the  nodes for sending requests to their chosen PNs and the second section is a scheduled section in which every PN has its own time-slot for transmission.
In MRC-based multi-hop broadcast, when a CN chooses multiple PNs, each of the selected PNs, if not having a CN already, needs to reserve a time-slot for its transmission in the scheduled section.
Moreover, each PN  needs to inform its neighboring nodes about the new CN that joined it.
 
\begin{figure} [!t]
   \centering
   \includegraphics[width=.6   \columnwidth]{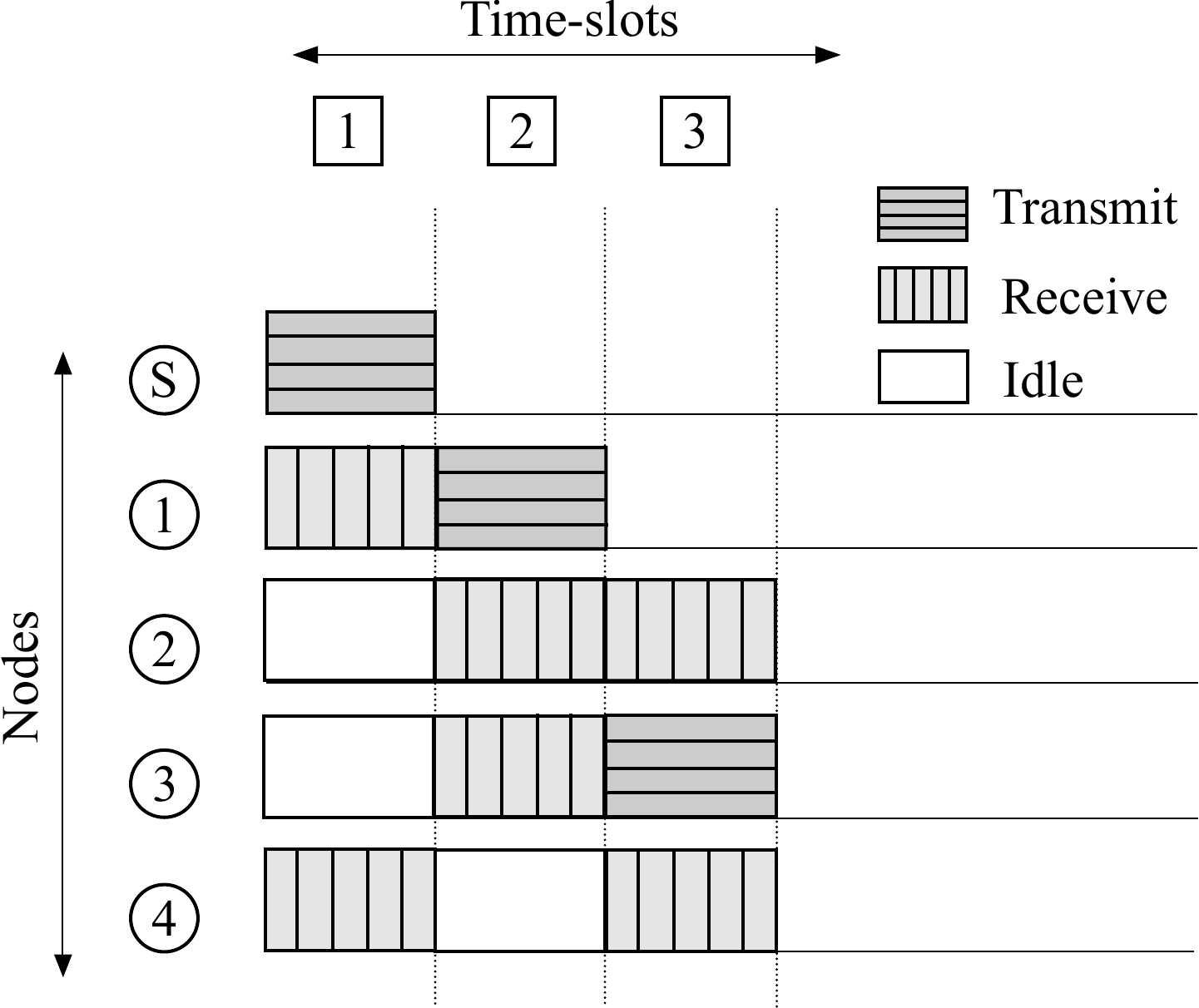}
   \caption{Transmission and reception in different time-slots in the scheduled section of the channel access.    }
   \label{fig:recp_model}
\end{figure} 
 
Since in MRC a node needs to receive the messages from its PNs in multiple time-slots, depending on the number of PNs that a CN chooses, the reception power  and consequently the energy it requires to consume over multiple time-slots  changes.
This issue is further illustrated in Fig. \ref{fig:recp_model} in which transmissions over the scheduled section of the channel access is shown.
In this Figure, for instance, node $\circled{1}$ receives its message from the source in time-slot \squared{1} and acts as a PN in the next time-slot.
Node  $\circled{2}$,  in order to receive the message,  accumulates the signals transmitted by nodes $\circled{1}$ and $\circled{3}$ in time-slots \squared{2} and \squared{3}, respectively.
In this example,  nodes $\circled{2}$ and $\circled{4}$ receive the message in two time-slots and the circuitry power they require for message reception is more than that required by nodes $\circled{1}$ and $\circled{3}$ that use only one time-slot.
Since every transmitting node in this network has its own time-slot for transmission and the receiving nodes merely receive the message during the time-slots that their respective PNs transmit, intra-network interference will be prevented in the network.
The nodes in this network need to be synchronized in time domain so that they access the channel at the correct point in time.
For clock synchronization, the clock of the source can be considered as the reference clock \cite{Wu_SPmag_11}.
Since the MRC at a CN is carried out over multiple time-slots, no phase synchronization is required among the PNs of the CN.
Note that collisions may occur during the network set-up in which the CNs find their respective PNs. 
In case of a collision, the nodes may back-off and access the channel at a later point in time\footnote{Please refer to \cite{Mahdi_thesis} for further discussion on the channel access scheme.}.

The reception power of a CN $i$ is calculated by 
\begin{equation}
\label{eq:mrc:prx_def}
P_i^\mathrm{Rx}( \bm{a} ) = W_i p_i^\mathrm{cr}.
\end{equation}
Hence, the \textit{total power} of node $j$ for both message transmission and reception is given by
\begin{equation}
\label{eq:mrc:ptot}
P_j^\mathrm{tot} \left( \bm{a} \right) = P_j^\mathrm{Rx}( \bm{a} ) +  P_j^\mathrm{Tx} ( \bm{a} ) 
\end{equation}

\subsection{Problem formulation}
\label{sec:mrc:prob_form}
Since the  transmission  time-slots have equal length, by omitting the duration of the time-slots, the energy consumption of a node for transmission/reception can be represented by the power required for transmission/reception.
Before defining the problems formally, we define the following terms.

\begin{definition} \label{def:mrc:net_power} 
\textbf{(Network power):} 
The network power is defined as the sum of the total power of the nodes as
\begin{equation}
\label{eq:mrc:p_tot_def}
P^\mathrm{tot}_\mathrm{net} \left(   \bm{a} \right)
:=
 \sum_{j \in  \mathcal{Q}} P_j^\mathrm{tot} ( \bm{a} ),
\end{equation}  
in which $P_j^\mathrm{tot} ( \bm{a} )$ is defined in \eqref{eq:mrc:ptot}.
We further define the network transmit power by $P^\mathrm{Tx}_\mathrm{net} \left(   \bm{a} \right) = \sum_{j \in  \mathcal{Q}} P_j^\mathrm{Tx} ( \bm{a} )$.
\end{definition} 

\begin{definition} \textbf{(Node's Cost):} 
We define by $C_i (\bm{p}_i^\mathrm{req})$ the total cost that a node $i$ pays to its chosen PNs as 
\begin{equation}
\label{eq:costdef_mrc}
C_i^f(\bm{a} )  = \sum_{j \in \mathcal{W}_i} c_{j,i}^f( \bm{a} ), 
\end{equation}
in which $c_{j,i}^f( p_{i,j}^\mathrm{req}  )$ is the cost paid \textit{from} node $i$ \textit{to} its PN $j$ under the cost sharing scheme $f$.
\end{definition}

\begin{definition} \label{def:social_cost} 
\textbf{(Social cost):} 
The  social cost  of a network formed by the nodes in $\mathcal{P}$ is defined as \cite{MAS08}
 \begin{equation}
 \label{eq:soc_cst_def} 
\mathrm{SC} \left(  \bm{a}  \right)
:=
\sum_{i \in  \mathcal{P}} C_i^f ( \bm{a}  ),  
\end{equation}  
in which $C_i (\bm{a}) $ is defined in \eqref{eq:costdef_mrc}.
\end{definition} 

\begin{definition} \label{def:bb_cost} 
\textbf{(Budget-balanced cost sharing scheme):}
A cost sharing scheme $f$ is budget balanced if 
\begin{equation}
\label{eq:BB_def}
\sum_{i \in \mathcal{M}_j} c_{j,i}^f ( \bm{a}  ) =  P_j^\mathrm{Tx} (\bm{a}).
\end{equation} 
\end{definition}
In other words, a cost sharing scheme is budget-balanced  if the summation of the costs assigned to the CNs of a PN is equal to the sum transmit power of the PN.

As discussed earlier, we consider two different scenarios in this network with the following objectives.
 
\textbf{Scenario 1 (Energy minimization)}
The objective in this scenario is formally defined as
\begin{subequations} \label{eq:MRC:opt:gob1_all} 
\begin{eqnarray}  \label{eq:MRC:gob1}
\mathrm{\textbf{P1:}}  \quad & \underset{ \{\bm{a}_i \}_{i \in \mathcal{P}} }{\text{minimize}}  &  
P^\mathrm{tot}_\mathrm{net} \left(\left\lbrace  \bm{a}_i \right\rbrace_{i \in \mathcal{P}} \right)  \qquad \qquad \\  
& \text{subject to:}   &  \gamma_{i \vert_\mathrm{MRC}} ( \bm{p}_i^\mathrm{req} ) \geq \gamma^\mathrm{th}  \label{cost:snr}  \\ 
& 						&  i \in \mathcal{P} , j \in \mathcal{A}_i \label{cost:i_j}. 
\end{eqnarray} 
\end{subequations}
According to the definition of $\mathcal{A}_i$ in \eqref{eq:mrc:action_set} and based on the definition of $h_i$ in \eqref{eq:hop_count_def}, it is straightforward to see that having the condition $j \in \mathcal{A}_i$ in \eqref{cost:i_j} is enough to ensure that the outcome of \textbf{P1} is a DAG.

\textbf{Scenario 2 (Social cost minimization)}
In this scenario, we aim to find the minimum cost required to be paid by the receiving nodes of the network for obtaining the message.
In this scenario,  we focus on the class of budget-balanced cost sharing schemes, cf. Definition~\ref{def:bb_cost}.
The network objective for the second scenario is defined as:
\begin{subequations} \label{eq:MRC:opt:gob2_all} 
\begin{eqnarray}  \label{eq:MRC:gob2}
\mathrm{\textbf{P2:}}  \quad & \underset{ \{\bm{a}_i \}_{i \in \mathcal{P}} } {\text{minimize}}  & \mathrm{SC} \left( \left\lbrace \bm{a}_i \right\rbrace_{i \in \mathcal{P}} \right)   \\
 & \text{subject to:}   &  \eqref{eq:BB_def}, \eqref{cost:snr}, \eqref{cost:i_j}.
\end{eqnarray} 
\end{subequations}
  
Table \ref{tab:notations} lists key notations used in this paper.
In the next section, we propose decentralized approaches for problems \textbf{P1} and \textbf{P2}.

\begin{table}  
	\centering
  	\caption{Key notations used in this paper.} 
\begin{center}
 \begin{tabular}{|p{1cm}||p{6.9cm}|} 
 \hline
 Notation & Definition  \\
 \hline
  \addlinespace[.5mm]
  \hline
$N$    & Number of receiving nodes \\ \hline
$\mathcal{P}$    & The set of all the nodes of the network except the source \\ \hline 
$\mathcal{Q}$    & The set of all the nodes of the network \\ \hline 
$\mathcal{W}_i$    & The set of PNs chosen by CN $i$ \\ \hline 
$\mathcal{M}_j$    & The set of CNs of PN $j$ \\ \hline 
$M_j$    & Number of CNs of PN $j$ \\ \hline
$W_i$    & Number of PNs of node $i$\\ \hline 
$W^\mathrm{max}$    & Maximum number of PNs allowed to be chosen by a CN\\ \hline 
$h_i$    & Hop-rank of node $i$	\\ \hline 
$\gamma_i$    & SNR of the signal received at CN $i$ \\ \hline 
$\gamma^\mathrm{th}$    & Minimum SNR required at a CN for decoding the signal \\ \hline 
$p_j^\mathrm{ct}$, $p_j^\mathrm{cr}$		& Circuitry power of node $j$ for transmission, reception.	\\ \hline
$p_j^\mathrm{Tx}$				& Radio-link power of node $j$ 			  								\\ \hline
$P_j^\mathrm{Tx}$				& Sum transmit power of node $j$: $ p_j^\mathrm{ct} + p_j^\mathrm{Tx}$	 	\\ \hline
$P_j^\mathrm{Rx}$				& Reception power of node $j$: $ W_j p_j^\mathrm{cr}$	 	\\ \hline
$P_j^\mathrm{tot}$				& Total power of node $j$ :	$P_j^\mathrm{Rx} + P_j^\mathrm{Tx}$		  	\\ \hline
$P_\mathrm{net}^\mathrm{Tx}$   & Network transmit power: $\sum_{j \in \mathcal{Q}} P_j^\mathrm{Tx}$  		\\ \hline
$P_\mathrm{net}^\mathrm{tot}$    & Network power: $\sum_{j \in \mathcal{Q}} P_j^\mathrm{tot}$   	\\ \hline 
$\bm{p}_i^\mathrm{req}$    & Vector of power requests of CN  $i$ from its chosen PNs\\ \hline 
$\bm{p}_{j}^\mathrm{rcv}$    & Power request vector received by PN $j$ \\ \hline 
$\bm{p}_{-i,j}^\mathrm{rcv}$    & Power request vector received by PN $j$ excluding the request of CN $i$ \\ \hline 
$\bm{a}_{i}$    & Action of node $i$ \\ \hline 
$\bm{a} $    & Set of actions of all the nodes of the network \\ \hline 
$\bm{a}_{-i}$    & Set of actions of all the nodes of the network except node $i$ \\ \hline 
$c_{j,i}^f$    & Cost paid by CN $i$ to PN $j$ under the sharing function $f$ 	\\ \hline 
$C_{i}^f$    & Total cost paid by CN $i$ to all its PNs \\ \hline 
\end{tabular}
\end{center}
\label{tab:notations}
\end{table}

\section{MRC-based decentralized multi-hop broadcast} 
\label{sec:MRCgame}
	\subsection{Game-theoretic model} 
	\label{sec:game_properties}
We design our decentralized approach via a non-cooperative CSG with the following properties:

\begin{itemize}

\item
\textbf{Players:}
The finite number of nodes in  $\mathcal{P}$.

\item
\textbf{Action:} Defined in \eqref{eq:mrc:decision} as a set of tuples that determines the PN set and (radio-link) power request set of the node, defied in \eqref{eq:mrc:pn} and \eqref{eq:mrc:pow_req}, respectively. 
For the PN set of CN $i$ we have $\mathcal{W}_i \in 2^{\mathcal{A}_i} \backslash \{ \varnothing \}$ and the joint PN set of the game is given by $\mathcal{W} = \bigtimes_{i \in \mathcal{P}} \mathcal{W}_i$, in which $ 2^{\mathcal{A}_i}$  and $\bigtimes$ represent  the  power-set of $\mathcal{A}_i$ and the Cartesian product, respectively.
We further define the joint request set of the game as $\mathcal{P}^\mathrm{req} = \bigtimes_{i \in \mathcal{P}} \bm{p}_i^\mathrm{req}$.

\item
\textbf{Cost function:} Assigns a real-valued cost to every node $i \in \mathcal{P}$ as $C_i(\bm{a}): \left( \mathcal{W} , \mathcal{P}^\mathrm{req} \right) \rightarrow \mathbb{R}^+$.
The cost function is defined in \eqref{eq:costdef_mrc}.

\end{itemize}

\begin{remark}
We employ the  MC cost sharing scheme for the first scenario.
We showed in \cite{Mousavi_TWC18} that the MC is simple and performs fairly well for network power minimization in multi-hop broadcast.
\end{remark}

\begin{definition} \textbf{(Marginal contribution (MC))} \label{def:MC}
The cost of node $i$, based on the MC, is defined as the power imposed by CN $i$ on the network, including the sum transmit power of the chosen PNs and  its own circuitry power, as
\begin{align}
\label{eq:mc_def}
C_{i}^\mathrm{MC}(\bm{a}) & =   P_i^\mathrm{Rx}(\bm{a}) + \sum_{j \in \mathcal{W}_i} c_{j,i}^\mathrm{MC}(\bm{p}_j^\mathrm{rcv})   \\
& = W_i p_i^\mathrm{cr} \quad + \sum_{j \in \mathcal{W}_i} P_j^\mathrm{Tx} (\bm{p}_j^\mathrm{rcv}) - P_j^\mathrm{Tx} (\bm{p}_{-i,j}^\mathrm{rcv}). \notag
\end{align}
with  $P_j^\mathrm{Tx} (\bm{p}_j^\mathrm{rcv})$	 defined in \eqref{eq:mrc:ptx_def}.
\end{definition}

\begin{remark}
In the second scenario where the CNs need to pay a price for receiving data, there are two points that have to be considered in defining a cost sharing function.  
Firstly, the cost sharing function must be budget-balanced.

With a budget-balanced scheme, as shown in \eqref{eq:BB_def}, a PN is paid by its CNs according to its sum transmit power.
Secondly, the cost sharing function must be fair.
Given a multicast by a PN,  according to \eqref{eq:p_radio}, each CN imposes a different level of radio-link power on the PN and thus, the cost that they need to pay to the PN must be shared among them in a fair manner.
For this scenario, we employ Shapley value (SV) which is budget-balanced and known as the fairest cost sharing scheme \cite{MAS08}.
\end{remark}

\begin{definition} \textbf{(Shapley value (SV))} \label{def:SV}
The SV of CN $i\in \mathcal{M}_j$ is defined by \cite{S53}
\begin{equation}
\label{eq:shapley_general}
c_{j,i}^\mathrm{SV} ( \bm{a} )  = 
\smashoperator[lr]{\sum_{\mathcal{S} \subseteq \mathcal{M}_j \setminus \{i\}} }
\frac{|\mathcal{S}|!\; (|\mathcal{M}_j|-|\mathcal{S}|-1)!}{|\mathcal{M}_j|!}(P_j^\mathrm{Tx}(\mathcal{S}\cup\{i\})-P_j^\mathrm{Tx}(\mathcal{S})).
\end{equation} 
\end{definition}

\begin{remark}
While the cost function of a node in the first scenario, defined in \eqref{eq:mc_def}, contains  the reception  power of a CN, the one used for the second scenario is merely defined according to the \textit{ sum transmit powers} of the PNs. 
The reason is that the goal in the first scenario is to minimize the network power and as far as the power of the network as a whole is concerned, the reception power needs to  be also included in the cost function of the node.
In contrast, in the second scenario we aim at minimizing the  paid cost.
In this case, we assume that a CN is not concerned about its reception power and merely aims at minimizing the cost it pays to its PNs.
\end{remark}

\begin{definition} \textbf{(Exact potential game)} \label{def:PotDef}
A game $G$ is an exact potential game \cite{Monderer96} if there exists a function $\Phi: \left( \mathcal{W} , \mathcal{P}^\mathrm{req} \right) \rightarrow \mathbb{R}$, called the potential function, such that $\forall i \in \mathcal{P}$,
$ \bm{a}, \bm{a}'  \in  \left( \mathcal{W} , \mathcal{P}^\mathrm{req} \right)$
\begin{equation} \label{eq:PotDef}
C_i(\bm{a}^\prime)  - C_i(\bm{a}) = \Phi(\bm{a}^\prime)   - \Phi(\bm{a}).
\end{equation}
\end{definition}

\begin{lemma} \label{lem:pot_opn}
The game $G$ without employing MRC, where every node is allowed to choose only one PN,  is a potential game for which at least one pure NE exists \cite{Mousavi_TWC18}.
The NE is reachable by using the best response dynamics in which every node iteratively chooses its PNs in a way to minimize its cost given the decision of other nodes.
\end{lemma}

\begin{theorem}
The game $G$ with MRC possesses an NE.
\end{theorem}
\begin{proof}
Using Lemma \ref{lem:pot_opn} and the definition of an exact potential game in Definition \ref{def:PotDef}, we show that our proposed game is also an an exact potential game.
Let $\underset{(j)} {\Delta^i} c_{i}$ and $\underset{(j)} {\Delta^i} \Phi$ denote the change in the cost of node $i$ and the potential function when $i$ changes its request from  PN $j$.
Being a potential game in case of having only one PN implies that $\underset{(j)} {\Delta^i} c_{i} = \underset{(j)} {\Delta^i} \Phi$.
Let $\mathcal{W}_i$ and $\mathcal{W}_i^\prime$ be the sets of old and new PNs of CN $i$ and $\underset{\mathcal{W}_i \rightarrow \mathcal{W}_i^\prime}{\Delta^i} C_i$ be the change in the cost of node $i$ when it changes its PNs from $\mathcal{W}_i$ to $\mathcal{W}_i^\prime$. 
According to  \eqref{eq:costdef_mrc}, the change in the cost of node $i$ can be written as
\begin{align}
\label{eq:mrc:ne_proof}
\underset{\mathcal{W}_i \rightarrow \mathcal{W}_i^\prime}{\Delta^i} C_i & = 
\underset{\mathcal{W}_i \rightarrow \mathcal{W}_i^\prime} {\Delta^i} \sum_{j \in \mathcal{W}_i \cup \mathcal{W}_i^\prime } c_{j,i}
\end{align}
Since the cost function in $G$, defined in \eqref{eq:costdef_mrc}, is linearly separable with respect to the cost paid by a CN to each of its chosen PNs,  we can write the right side of \eqref{eq:mrc:ne_proof} as
\begin{align}
\label{eq:mrc:ne_proof_2}
\underset{\mathcal{W}_i \rightarrow \mathcal{W}_i^\prime} {\Delta^i} \sum_{j \in \mathcal{W}_i \cup \mathcal{W}_i^\prime } c_{j,i} 
&  = \sum_{j \in \mathcal{W}_i \cup \mathcal{W}_i^\prime } \underset{(j)} {\Delta^i} c_{i}  
 = \sum_{j \in \mathcal{W}_i \cup \mathcal{W}_i^\prime } \underset{(j)} {\Delta^i} \Phi \notag \\
& = \underset{(j)} {\Delta^i} \sum_{j \in \mathcal{W}_i \cup \mathcal{W}_i^\prime } \Phi  
 = \Delta \Phi^\prime, 
\end{align}
which shows that the game is still an exact potential game.
Hence, at least one NE exists for the game which can be obtained using the best response dynamics.
\end{proof}

\begin{remark}\label{rmk:complexity}
Although reaching an NE in a finite number of iterations is guaranteed by using the best response dynamics, the convergence time is exponential in the worst case.
It has been shown in \cite{Durand_agt_springer_16} that the average convergence time of the best response dynamics is $e^\zeta N + O(N)$ in which $e$ and $\zeta$ are the Euler's number and the Euler's constant, respectively. Such a convergence time is acceptable for practical scenarios.
\end{remark}
 
	\subsection{Scenario 1: Minimum-power data dissemination} 
	\label{sec:MRCMC}

In this section, we propose an MILP for the problem in \eqref{eq:MRC:gob1} using the MC, aiming at network power minimization.
 According to \eqref{eq:mc_def}, $c_{j,i}^\mathrm{MC}( \bm{a} )$ can be broken down as  \eqref{eq:mrc:mc_breakdown}.
 
\begin{figure*}[t]
\begin{equation}
\label{eq:mrc:mc_breakdown}
c_{j,i}^\mathrm{MC}( \bm{a} )  = 
\begin{cases}
 p_j^\mathrm{ct} + p_{i,j}^\mathrm{req}, &\text{if } \mathds{1}_j(\mathcal{M}_j\backslash \{i\}) = 0 \\ 
 p_{i,j}^\mathrm{req} - p_j^\mathrm{Tx} ( \bm{p}_{-i,j}^\mathrm{rcv} ), 
&\text{if } \mathds{1}_j(\mathcal{M}_j\backslash \{i\}) = 1  , p_{i,j}^\mathrm{req} = \underset{h \in \mathcal{M}_j}{\max} \{ p_{h,j}^\mathrm{req} \} \\
  0 & \text{if } \mathds{1}_j(\mathcal{M}_j\backslash \{i\}) = 1  , p_{i,j}^\mathrm{req} \neq \underset{h \in \mathcal{M}_j}{\max}\{ p_{h,j}^\mathrm{req} \}.
\end{cases}
\end{equation}
\hrule
\end{figure*}

As can be seen from \eqref{eq:mrc:mc_breakdown}, when $\mathds{1}_j(\mathcal{M}_j \{i\}) = 1$, the cost function $c_{j,i}^\mathrm{MC}( \bm{a} )$ is a piece-wise linear function.
Given the piece-wise linearity of the cost function, we propose the MILP problem shown in  \eqref{eq:MRC:opt:mc:all} for decision making at every node $i \in \mathcal{P}$.
In the MILP, given in \eqref{eq:MRC:opt:mc:all}, $\bm{w}_i$ is a binary vector of length $|\mathcal{A}_i|$ defined as $\bm{w}_i = [w_{i,j}|j \in \mathcal{A}_i ]$ where $w_{i,j} = 1$  if node $i$ chooses node  $j \in \mathcal{A}_j$ as its PN.
 
\begin{figure*}[t]
\begin{subequations} \label{eq:MRC:opt:mc:all} 
\begin{alignat}{6}  \label{eq:MRC:opt:mc} 
& \argmin_{ \bm{p}_{i}^\mathrm{req}, \bm{t}_i,  \bm{w}_i} \quad & & \sum_{j\in \mathcal{A}_{i}} w_{i,j} p_i^\mathrm{cr}  +  t_{i,j} , &  \forall i \in \mathcal{P}  \\
& \text{subject to }    & &  \notag \\
&&& w_{i,j} p_j^\mathrm{min}  \leq p^\mathrm{req}_{i,j} \leq w_{i,j} p_j^\mathrm{max}, & \forall j \in \mathcal{A}_i \label{eq:mrc:mc:pdelta}\\
 &&& \sum_{j\in\mathcal{W}_i} \frac{  p_{i,j}^\mathrm{req} g_{i,j}} { \sigma^2} = \gamma^\mathrm{th} & \forall j \in \mathcal{A}_i \label{eq:mrc:mc:snr_cond} \\	
 &&& p_{i,j}^\mathrm{req} - p_j^\mathrm{Tx} ( \bm{p}_{-i,j}^\mathrm{rcv} )  + (1 - \mathds{1}_j(\mathcal{M}_j\backslash \{i\}) ) w_{i,j} p_j^\mathrm{ct} \leq t_{i,j} \quad &  \forall j \in \mathcal{A}_i \label{eq:mrc:mc:t_piecewise}\\
&&& \sum_{j \in \mathcal{A}_i} w_{i,j} \leq W^\mathrm{max}  &  \forall j \in \mathcal{A}_i \label{eq:mrc:mc:maxPN}\\
&&& p^\mathrm{req}_{i,j}, t_{i,j} \in \mathbb{R}, s_{i,j}, w_{i,j} \in \{0,1\}   &   \forall j \in \mathcal{A}_i  
\end{alignat}	
\end{subequations} 
\hrule
\end{figure*}

We define   $t_{i,j}$ as an auxiliary variable which is used in both \eqref{eq:MRC:opt:mc} and \eqref{eq:mrc:mc:t_piecewise}.
$w_{i,j} p_i^\mathrm{cr}$  in \eqref{eq:MRC:opt:mc} represents  the power imposed on the network by node $i$ due to its signal reception power.
Moreover, $t_{i,j}$ captures the sum transmit power of node $i$ via \eqref{eq:MRC:opt:mc} and \eqref{eq:mrc:mc:t_piecewise}.
In \eqref{eq:mrc:mc:pdelta}, $p_j^\mathrm{min}$ is the minimum radio-link power of a transmitter in transmit mode.
\eqref{eq:mrc:mc:snr_cond} represents the minimum-SNR condition for signal reception.
Finally, \eqref{eq:mrc:mc:maxPN} can restrict the number of PNs that a CN can choose if the system designer sets such a restriction.

	\subsection{Scenario 2: Minimum-cost data dissemination} 
	\label{sec:MRCSV}
Despite the complexity of the SV in  \eqref{eq:shapley_general}, we will show that it can be represented by a 
piece-wise linear function.

\begin{lemma} \label{lem:sv_with_pc}
$c_{j,i}^\mathrm{SV} ( \bm{a} )$ in \eqref{eq:shapley_general} can be written as
\begin{equation}
\label{eq:sv_with_pc}
c_{j,i}^\mathrm{SV} ( \bm{a} ) = \frac{p_j^\mathrm{ct}}{M_j} +  \sum\limits_{n=1}^{i}\frac{p_{n,j}^{\mathrm{req}} - p_{n-1,j}^{\mathrm{req}}}{M_j+1-n}.
\end{equation}
\end{lemma}

\begin{proof}
Using \eqref{eq:mrc:ptx_def}, the sum transmit power of a PN $j$ is composed of two parts; $p_j^\mathrm{ct}$ as the fixed circuitry  power and  the the radio link power, i.e., $p_j^\mathrm{Tx}$.
Since these two parts are independent and since the SV satisfies the additivity axiom \cite[Ch.~12]{MAS08}, the cost of a node $i  \in \mathcal{M}_j$ can be separated as 
\begin{equation}
\label{eq:sv_expand}
c_{j,i}^\mathrm{SV} (  \bm{a}  )  = c_{j,i}^\mathrm{SV} ( p_j^\mathrm{ct} + \bm{p}_{j}^\mathrm{rcv}  )  = c_{j,i}^\mathrm{SV} (p_j^\mathrm{ct})  + c_{j,i}^\mathrm{SV} ( p_{i,j}^\mathrm{req}, \bm{p}_{-i,j}^\mathrm{rcv}  ) 
\end{equation}
in which the cost related to the circutiry power is equally shared among the CNs  as  $c_{j,i}^\mathrm{SV} (p_j^\mathrm{ct}) = p_j^\mathrm{ct}/M_j$ \cite{Mousavi_TWC18}.
Moreover, the cost share regarding the requests of the CNs
by assuming that they can be sorted as 
\begin{equation}
\label{eq:pow_sort}
0 = p_{0,j}^\mathrm{req} \leq   p_{1,j}^\mathrm{req} \leq \dots \leq p_{n,j}^\mathrm{req} \leq p_{i,j}^\mathrm{req} \leq p_{n+2,j}^\mathrm{req} \leq \dots \leq p_{M_j,j}^\mathrm{req} 
\end{equation}
 is given by \cite{LO73} \cite{SA11}
\begin{equation}
\label{eq:sv_radio}
c_{j,i}^\mathrm{SV} ( p_{i,j}^\mathrm{req}, \bm{p}_{-i,j}^\mathrm{rcv}  )  = \sum\limits_{k=1}^{i}\frac{p_{k,j}^{\mathrm{req}} - p_{k-1,j}^{\mathrm{req}}}{M_j+1-k}.
\end{equation}
\end{proof}

\begin{lemma} \label{lem:sv_piecewise}
Suppose that the requests received by PN $j$ can be sorted as \eqref{eq:pow_sort}.
$c_{j,i}^\mathrm{SV} ( p_{i,j}^\mathrm{req}, \bm{p}_{-i,j}^\mathrm{rcv}  )$  in \eqref{eq:sv_expand}  can be modeled by a piecewise-linear, increasing function as
\begin{align}
\label{eq:sv_piece}
c^\mathrm{SV}_{j,i} &(p_{i,j}^\mathrm{req},\bm{p}_{-i,j}^\mathrm{rcv}) 
=
  \frac{p_{i,j}^\mathrm{req}}{M_j-n} 
+  \smashoperator[lr]{ \sum\limits_{k=1}^{n \geq 1} } 
 \left( \frac{ - p_{k,j}^\mathrm{req} } { (M_j-k ) (M_j-k+1)} \right).
\end{align}
 \end{lemma}
 \begin{proof}
 See Appendix \ref{app:a}.
 \end{proof}

\begin{theorem} \label{th:sv_p-wise}
Let $\mathcal{M}_{j\backslash i } := \mathcal{M}_j \backslash \{i\}$ be the set of CNs of PN $j$ without node $i$  with the number  $M_{j\backslash i} $ of CNs.
Let $p_{i,j}^\mathrm{req}$ be the $(n+1)$-th smallest request among the CNs of PN $j$ as \eqref{eq:pow_sort}.
Then, the cost of a node $i$ in \eqref{eq:sv_expand} if it joins PN $j$ according to the SV is obtained by  
\begin{equation}
\label{eq:mrc:SV_finform}
c_{j,i}^\mathrm{SV} (  \bm{a} ) =  m_i(n) p_{i,j}^\mathrm{req}  + y_i(n, \bm{p}_{-i,j}^\mathrm{rcv})
\end{equation}
in which   
\begin{equation}
 \label{eq:a_n}
m_i(n) = \frac{1}{M_{j\backslash i}+1-n}
\end{equation}
and 
\begin{align}
 \label{eq:b_n}  
y_i(n, \bm{p}_{-i,j}^\mathrm{rcv})& =  \frac{p_j^\mathrm{ct}}{M_{j\backslash i} +1} +  \notag \\
 &   \smashoperator[lr]{ \sum\limits_{k=1}^{n \geq 1 } } \left( \frac{ - p_{k,j}^\mathrm{req} } { (M_{j\backslash i}-k+1) (M_{j\backslash i}-k+2)} \right).
\end{align}
\end{theorem}

\begin{proof}
It follows directly from Lemma \ref{lem:sv_with_pc} and Lemma \ref{lem:sv_piecewise}.
\end{proof}

\begin{corollary}
 The optimum request vector of node $i$, i.e., $\bm{p}_i^\mathrm{req}$ can be obtained by solving an MILP.
\end{corollary}
 
 The optimal decision of node $i$ can be obtained by the MILP shown in \eqref{eq:mrc:sv:milp:all}.
Notice that  the parameters used in the MILP of  \eqref{eq:mrc:sv:milp:all} are like the ones used for the MC scheme in \eqref{eq:MRC:opt:mc:all}.

\begin{figure*}[t]
\begin{subequations}
\label{eq:mrc:sv:milp:all}
\begin{alignat}{6}  \label{eq:mrc:sv:milp}
& \argmin_{ \bm{p}_{i}^\mathrm{req}, \bm{t}_i, \bm{s}_i, \bm{w}_i} \quad & & \sum_{j\in \mathcal{A}_{i}}   t_{i,j} , &&  \forall i \in \mathcal{P}  \\
& \text{subject to: }    & &  \notag \\
&&& w_{i,j} p_j^\mathrm{min}  \leq p^\mathrm{req}_{i,j} \leq w_{i,j} p_j^\mathrm{max}, && \forall j \in \mathcal{A}_i \label{eq:pdelta}\\
 &&& \sum_{j\in\mathcal{W}_i} \frac{  p_{i,j}^\mathrm{req} g_{i,j}} { \sigma^2} = \gamma^\mathrm{th} && \forall j \in \mathcal{A}_i \label{eq:snr_cond} \\	
 &&& m_i(n) p_{i,j}^\mathrm{req}  + y_i(n, \bm{p}_{-i,j}^\mathrm{rcv}) \leq t_{i,j}\quad &&     \forall j \in \mathcal{A}_i,  0 \leq n \leq M_j   \label{eq:t_piecewise}\\
&&& \sum_{j \in \mathcal{A}_i} w_{i,j} \leq W^\mathrm{max}  &&  \forall j \in \mathcal{A}_i \label{eq:maxPN}\\
&&& p^\mathrm{req}_{i,j}, t_{i,j} \in \mathbb{R}, s_{i,j}, w_{i,j} \in \{0,1\}   &&   \forall j \in \mathcal{A}_i
\end{alignat}
\end{subequations}
\hrule
\end{figure*}

\section{MRC-based centralized approach with MILP}
\label{sec:MRCMILP}

In the previous section, we discussed decision-making  with the MC and the SV  schemes.
In this section, we find the global optimum for both the network power and the social cost minimization problems.

\begin{theorem}
\label{th:pEN_pSC}
Let $\{ \overset{*}{ \bm{a}_i } \}_{i \in \mathcal{P}}^\mathrm{PTx}$ and $\{ \overset{*}{ \bm{a}_i } \}_{i \in \mathcal{P}}^\mathrm{SC}$ be the set of optimum action profiles corresponding to the network's minimum \textit{ transmit power} (see Table \ref{tab:notations}) and the minimum social cost (with a budget-balanced cost sharing scheme), respectively.
We always have $\{ \overset{*}{ \bm{a}_i } \}_{i \in \mathcal{P}}^\mathrm{PTx} = \{ \overset{*}{ \bm{a}_i } \}_{i \in \mathcal{P}}^\mathrm{SC}$.
\end{theorem}
\begin{IEEEproof}
Theorem \ref{th:pEN_pSC} states that any  action profile that minimizes the network's \textit{transmit power}, minimizes the social cost as well.
Using the definition of a budget-balanced cost sharing scheme, for every PN $j \in \mathcal{N}$ we have  
\begin{equation}
\sum_{i \in \mathcal{M}_j} c_{j,i}^\mathrm{BB} ( \bm{a} ) = P_j^\mathrm{Tx}( \bm{a} ).
\end{equation}
By taking a summation over all the nodes of the network which can act as PNs, we get
\begin{equation}
 \label{eq:SumCost}
 \sum_{j \in \mathcal{Q}} \sum_{i \in \mathcal{M}_j} c_{j,i}^\mathrm{BB}( \bm{a} )  =  \sum_{j \in \mathcal{Q}}  P_j^\mathrm{Tx}( \bm{a} ).
\end{equation}
Note that  the payment received by a PN is equal to the cost paid by its CNs.
Therefore, we can replace the left side of \eqref{eq:SumCost}, which is the total payment received by the PNs in the network, with the total cost paid by the CNs.
Hence, the left side of \eqref{eq:SumCost}, using  \eqref{eq:costdef_mrc}, is equivalent to
\begin{align}
 \label{eq:leftof}
 \sum_{j \in \mathcal{Q}} \sum_{i \in \mathcal{M}_j} c_{j,i}^\mathrm{BB}( \bm{a} )&  = \sum_{i \in \mathcal{P}} \sum_{j \in \mathcal{W}_i} c_{j,i}^\mathrm{BB}(\bm{a})   \notag \\
 & =   \sum_{i \in \mathcal{P}} C_i(\bm{a})
  =   \mathrm{SC} (\bm{a}) 
\end{align}
By comparing \eqref{eq:SumCost} and \eqref{eq:leftof}, we find that with a budget-balanced cost sharing scheme, the action profile that minimizes the social cost also minimizes the network transmit power.
\end{IEEEproof} 
 
\begin{corollary}
The MILP that finds the optimum configuration for P1 in \eqref{eq:MRC:opt:gob1_all} can also be used for P2 in \eqref{eq:MRC:opt:gob2_all} if $p_j^\mathrm{cr} = 0$.
\end{corollary}

Before providing the MILP, we define the following.

\begin{itemize}

\item
$\tilde{S}$: The maximum number of time-slots used for message dissemination in the network. 
To avoid notational confusion,   between the source and the time-slot, we show the source node in this formulation by $\underline{\mathrm{S}}$ and $(s)$ represents  time-slot $s$ such that $1 \leq  s  \leq \tilde{S}$.

\item
$\bm{P}^\mathrm{Tx}$ (radio-link power matrix): 
A $(N+1) \times \tilde{S}$ matrix with $\bm{P} = [\bm{p}_{(1)}^\mathrm{Tx},  \dots, \bm{p}_{(\tilde{S})}^\mathrm{Tx}]$ in which $\bm{p}_{(s)} = [p_{\mathrm{\underline{S}},(s)}^\mathrm{Tx}, p_{1,(s)}^\mathrm{Tx}, \dots,  p_{N,(s)}^\mathrm{Tx}]^\intercal$ is a column vector and $p_{j,(s)}^\mathrm{Tx}$ shows the radio-link power of node $j$ in time-slot $s$.

\item
$\bm{T}$ (transmission matrix): 
A $(N+1) \times \tilde{S}$ binary matrix with $\bm{T} = [\bm{t}_\mathrm{\underline{S}}, \bm{t}_1,  \dots, \bm{t}_N]^\intercal$ in which $\bm{t}_j = [t_{j,(1)}, t_{j,(2)}, \dots,  t_{j,(\tilde{S})}]$ is a row vector of size $1\times \tilde{S}$ and $t_{j,(s)} = 1$ if node $j$ transmits in time-slot $s$.

\item
$\bm{R}$ (reception matrix): 
A $N  \times \tilde{S}$ binary matrix with $\bm{R} = [\bm{r}_1,  \dots, \bm{r}_N]^\intercal$ in which $\bm{r}_j = [r_{i,(1)}, r_{i,(2)}, \dots,  r_{i,(\tilde{S})}]$ is a row vector of size $1\times \tilde{S}$ and $r_{i,s} = 1$ if node $i$ receives the message from a transmitting node that transmits in time-slot $s$.

\item
$\bm{G}$ (channel gain matrix): 
A $N \times (N+1)$ matrix with $\bm{G} := [\bm{g}_1^\intercal, \dots, \bm{g}_N^\intercal]^\intercal$ in which $\bm{g}_i = [g_{i,\mathrm{\underline{S}}}, g_{i, 1} \dots, g_{i,N}] $ is  a $ 1 \times N+1$ row vector and $g_{i,j}$ is the channel gain between transmitter $j$ and receiver $i$.
We set $g_{i,j} = 0$ if $j \notin \mathcal{A}_i$.

\item
$\bm{\Gamma}$ (SNR matrix): 
A $N \times \tilde{S}$ matrix with $\bm{\Gamma} := [\bm{\gamma}_1, \dots, \bm{\gamma}_N]^\intercal$ in which $\bm{\gamma}_i = [\gamma_{i,\mathrm{\underline{S}}}, \gamma_{i, 1} \dots, \gamma_{i,N}] $ is  a $ 1 \times \tilde{S}$ row vector and $\gamma_{i,(s)}$ is the SNR received by node $i$ in time-slot $s$.

\item
$\bm{1}_N$: An all 1 vector of size $N\times 1$.
 
\item
$ \mathds{I}_{N(i)} $: A binary vector of size $N \times 1$ with all its elements equal to 1 except the $i$-th element.
\end{itemize}

Before presenting the optimization problem, we provide the following lemma.
\begin{lemma} \textbf{(Big M method)}:
\label{lem:bigM}
Let $v_1, v_2 \in \mathbb{R}_{\geq 0} $ and $b\in \{0,1\}$ all be the variables of an optimization problem.
The following non-linear constraint 
\begin{equation}
v_1 = v_2 b 
\end{equation}

can be linearized by the following set of constraints 
\begin{align}
& v_1 \leq v_2 \label{lem:bigM2} \\
& v_1 \leq \mathds{M} b \label{lem:bigM1}\\
& v_1 \geq v_2 - \mathds{M} (1-b) \label{lem:bigM3}\\
& v_1 \geq 0   \label{lem:bigM4}
\end{align}
in which $\mathds{M}$ is a sufficiently large number such that $v_2 \leq \mathds{M}$.
\end{lemma} 

\begin{remark}
According to Lemma \ref{lem:bigM}, when $b=1$, based on the constraints \eqref{lem:bigM2} and \eqref{lem:bigM3}, $v_2$ limits $v_1$ from both upper and lower sides as  
$v_2 \leq v_1 \leq v_2$. 
Hence, $v_1 = v_2$ if $b =1$.
Likewise, one can observe that when $b=0$, according to \eqref{lem:bigM1} and \eqref{lem:bigM3} we have $v_2 - \mathds{M} \leq v_1 \leq 0$.
Further, based on the constraint \eqref{lem:bigM4} together with  \eqref{lem:bigM1} and \eqref{lem:bigM3} we finally get $0 \leq v_1 \leq 0$ which results in $v_1 = 0$.
\end{remark}

The MILP for the MRC-based multi-hop broadcast is presented  in \eqref{eq:milp_mrc_all}. 
Recall that, when the network objective is social cost minimization,  we set $p_j^\mathrm{cr} = 0$ in formulation \eqref{eq:milp_mrc}.
 
\begin{figure*}[t]
\begin{subequations} \label{eq:milp_mrc_all}
\begin{alignat}{6}  \label{eq:milp_mrc}
& \min_{ \bm{P}, \bm{T}, \bm{R}, \bm{\Gamma}} \quad & &  \sum_{s = 1}^{\tilde{S}} \Big( \sum_{j \in \mathcal Q}  p_{j,(s)}^\mathrm{Tx} +  p_j^\mathrm{ct} t_{j,(s)} + \sum_{i \in \mathcal P} p_j^\mathrm{cr}  r_{i,(s)} \Big), \quad  && \\
& \text{s. t.: }    & &  \notag \\
&&&  t_{j,(s)} p_j^\mathrm{min}  \leq p_{j,(s)}^\mathrm{Tx}  \leq   t_{j,(s)} p_j^\mathrm{max}, 	&& \forall j \in \mathcal{Q}, 1\leq s \leq \tilde{S} 	\label{eq:ILP:pdelta}\\
&&& \bm{t}_j \bm{1}_{\tilde{S}} \leq 1, 													&& \forall j \in \mathcal{Q} 					\label{eq:ILP:tx}\\
&&& \bm{r}_i \bm{1}_{\tilde{S}} \geq 1, 													&& \forall i \in \mathcal{P} 					\label{eq:ILP:rx}\\
&&& \sum_{j \in \mathcal{Q}} t_{j,(s)} \leq 1 ,  											&& \forall  j \in \mathcal{Q}, 1 \leq s \leq \tilde{S} 	\label{eq:ILP:singletx}\\
&&& r_{i,(s)} \leq \sum_{j \in \mathcal{Q}} t_{j,(s)} ,  									&& \forall  i \in \mathcal{P}, j \in \mathcal{Q}, 1 \leq s \leq \tilde{S} 	\label{eq:ILP:rx_tx}\\
&&& r_{i,(s)} \leq 1 - \sum_{s^\prime=1}^{s} t_{i, (s^\prime)} , 					&& \forall j \in \mathcal{Q}, 1 \leq s \leq \tilde{S}	\label{eq:ILP:no_rx_aft_tx}\\
&&& \hat{\gamma}_{i,(s)} = \bm{1}_{N+1}^\intercal  \left[ \bm{p}_{(s)} \odot \mathds{I}_{N+1(i)}  \odot \bm{g}_i^\intercal \right] /\gamma^\mathrm{th} \sigma^2, 	\qquad								&& \forall j \in \mathcal{Q}, 1 \leq s \leq \tilde{S} 	\label{eq:ILP:gamma_norm_calc}\\
&&&  \gamma_{i,(s)} \leq \hat{\gamma}_{i,(s)} , 									&& \forall i \in \mathcal{P}, 1 \leq s \leq \tilde{S} 	\label{eq:ILP:gamma_big_m_0}\\
&&&  \gamma_{i,(s)} \geq \hat{\gamma}_{i,(s)} - \mathds{M} (1-r_{i,(s)}) , 			&& \forall i \in \mathcal{P}, 1 \leq s \leq \tilde{S} 	\label{eq:ILP:gamma_big_m_1}\\
&&&  \gamma_{i,(s)} \leq \mathds{M} r_{i,(s)} , 									&& \forall i \in \mathcal{P}, 1 \leq s \leq \tilde{S} 	\label{eq:ILP:gamma_big_m_2}\\
&&& \bm{\gamma}_i \bm{1}_{\tilde{S}} \geq 1, 												&& \forall j \in \mathcal{P} 					\label{eq:ILP:snr}\\
&&&  t_{j,(s)} \leq \sum_{s^\prime = 1}^{s-1} \gamma_{j,(s^\prime)},				&& \forall j \in \mathcal{P}, 2 \leq s \leq \tilde{S} 	\label{eq:ILP:tx_gamma}\\
&&&  t_{\underline{\mathrm{S}},1} = 1 && \\
&&& p_{j,(s)}^\mathrm{Tx}, \gamma_{i,(s)} \in \mathbb{R}_{\geq 0}, t_{j,(s)}, r_{i,(s)} \in \{0,1\},   && \forall i\in\mathcal{P}, j \in \mathcal{Q}, 1\leq s\leq \tilde{S}    
\end{alignat}
\end{subequations}
\hrule
\end{figure*}

The constraint in  \eqref{eq:ILP:pdelta} indicates the radio-link power constraint.
Based on our assumption, every node can transmit in one time-slot.
This property has been captured by \eqref{eq:ILP:tx}.
The condition  in \eqref{eq:ILP:singletx} indicates that there exists at most one transmission per time-slot.
Further, every node receives the message at least in one time slot as shown in \eqref{eq:ILP:rx}.
The constraint in \eqref{eq:ILP:rx_tx} is due to the fact that a reception occurs in a time-slot if there is at least one transmission.
The constraint \eqref{eq:ILP:no_rx_aft_tx} indicates that a node $i \in \mathcal{P}$ does not receive the message if it has already transmitted it. 
In fact, $r_{i,(s)} = 0$ if node $i$ transmits the messages in one of the previous slots $1 \leq s^\prime \leq s$.
The expression in \eqref{eq:ILP:gamma_norm_calc}, in which $\odot$ shows the element-wise product, calculates the normalized SNR of the signal received at user $i$ in time-slot $s$.
More precisely, given the radio-link powers of the users at time-slot $s$ in $\bm{p}_s$, the expression 
$\bm{p}_s \odot \mathds{I}_{N+1(i)}  \odot \bm{g}_i^\intercal$
gives a vector whose elements are the SNR of the signal received by node $i$ from each of the transmitters $j \in  \mathcal{Q}$ in time-slot $s$.
Recall that $\mathds{I}_{N+1(i)} $ is an all-one column vector of length $N+1$ with the $i$-th element equal to zero.
This helps us to eliminate the SNR received by node $i$ due to its own transmission.
In \eqref{eq:ILP:gamma_norm_calc},  $\bm{1}_{N+1}^\intercal  \left[ \bm{p}_{(s)} \odot \mathds{I}_{N+1(i)}  \odot \bm{g}_i^\intercal \right]$ gives the SNR received by node $i$ which we normalize it to $\gamma^\mathrm{th} \sigma^2$ and represent it by $\hat{\gamma}_{i,(s)}$ with $\hat{\gamma}_{i,(s)} \leq 1$.

Although \eqref{eq:ILP:gamma_norm_calc} determines the SNR available at node $i$ in time-slot $s$, the actual SNR received by node $i$ depends on whether  node $i$ receives the message in this time-slot.
Node $i$ uses the signals transmitted in time-slot $s$ if $r_{i,(s)} = 1$.
The constraints in \eqref{eq:ILP:gamma_big_m_0}, \eqref{eq:ILP:gamma_big_m_1} and \eqref{eq:ILP:gamma_big_m_2} are used based on the big M  method, discussed in Lemma \ref{lem:bigM}.
They have been employed here to find out if node $i$ should receive the message in time-slot $s$.
More precisely, they linearize  the following constraint 
\begin{equation}
\label{eq:linear_non}
\gamma_{i,(s)} = r_{i,(s)} \hat{\gamma}_{i,(s)}
\end{equation}
in which $\hat{\gamma}_{i,(s)}$ is a function of the continuous variable $p_{i,(s)}^\mathrm{Tx}$ that makes the right side of \eqref{eq:linear_non}  non-linear.
The constraint in \eqref{eq:ILP:snr} indicates that the aggregated SNR obtained by every node $i \in \mathcal{P}$ must be higher than the minimum SNR.
Every node $i \in \mathcal{P}$ can transmit the message in time-slot $s$ if and only if it receives the message with minimum SNR over the previous time-slots.
The only exception is the source for which we always have $t_{\underline{\mathrm{S}},1} = 1$.

\section{Performance analysis}
\label{sec:simul_multiPN}

\begin{table}  
	\centering
  	\caption{Main parameters used for simulation} 
\begin{center}
 \begin{tabular}{|c||l|} 
 \hline
 Notation & value  \\
 \hline
  \addlinespace[.5mm]
  \hline
Number $N$ of nodes		& 10, 15, 20, 25	\\ \hline 
Network size			& 250m $\times$ 250m	\\ \hline 

Circuitry power, $p^\mathrm{c}$	& 1, 10, 100 mW	\\ \hline
Channel model						& Path-loss	\\ \hline
Path-loss coefficient, $\alpha$	& 3	\\ \hline
Noise power, $\sigma^2$			& -90 dBm	\\ \hline 
Signal wave-length, $\lambda$		& 0.125 m 	\\ \hline
 \end{tabular}
\end{center}
\label{tab:simul}
\end{table}

\subsection{Simulation setup}
An area of size 250m$\times$250m   is considered in which the nodes are randomly distributed.
The simulation results are  based on the Monte-Carlo method and in each simulation run, one of the nodes in the network is randomly chosen as the source.
The total number of nodes varies between 10 and 25.
We assume that the transmit and receive circuitry powers of the nodes in the network are equal, i.e.,  $p_j^\mathrm{cr} = p_j^\mathrm{ct} = p^c   $  \cite{Cui_TWC05} and consider  three values for circuitry powers, $p^c \in \{1, 10, 100\}  \text{ mW}$.
The low circuitry power case is suitable for low-power IoT applications while the high circuitry power can model conventional wireless transmitters \cite{Heinzelman_2000, wang_SECON06}.
The noise power, represented by $\sigma^2$ in \eqref{eq:snrdef}, is set to -90 dBm.

The channel is based on the path-loss model.  
Let  $l_{i,j}$ and $l_0$ be the distance between nodes $i$ and $j$ and a reference distance, respectively. Then, by considering  $\alpha$ as the path loss exponent and $\lambda$ as the signal wavelength, the power gain of the channel between nodes $i$ and $j$ is defined as 
\begin{equation}
 \label{eq:ChModel}
 g_{i,j} = \left(\frac{\lambda}{4 \pi l_0 } \right)^2 \left(\frac{l_0}{l_{i,j}}\right)^\alpha .
\end{equation}
For simulation, we set $\lambda=0.125$m, $l_0 = 1$m and  $\alpha=3$.

The results  are normalized to the value $v =\tilde{p}^\mathrm{Tx} + \tilde{p}^\mathrm{c}$ in which $\tilde{p}^\mathrm{Tx}$ and $ \tilde{p}^\mathrm{c}$ represent normalization reference values for the radio-link power and the circuitry power, respectively.
More precisely, the normalized network power and the normalized social cost are defined as
\begin{equation}
\label{eq:net_pow_norm}
\overbar{P}^\mathrm{tot}_\mathrm{net} (\bm{a}) = \frac{\sum_{j \in \mathcal{Q}} P_j^\mathrm{tot}  (\bm{a}) }{v}
, \quad
\overbar{\mathrm{SC}} (\bm{a}) = \frac{\sum_{i \in \mathcal{P}} C_i (\bm{a}) }{v} ,
\end{equation}
respectively, where $\bar{P}^\mathrm{tot}_\mathrm{net} (\bm{a})$ is defined in \eqref{eq:mrc:ptot} and $C_i(\bm{a}) = \sum_{j\in\mathcal{W}_i} c_{j,i}^\mathrm{SV}$ in which $c_{j,i}^\mathrm{SV} $ is defined in \eqref{eq:sv_with_pc}.
We set $\tilde{p}^\mathrm{Tx} = 200$ mW and $ \tilde{p}^\mathrm{c} = 10$ mW.
Moreover, we do not set any limitation on the number of PNs that a CN is allowed to select. 
The simulation has been carried out in MATLAB  and the proposed MILPs are solved using CVX \cite{cvx} and Gurobi\footnote{https://www.gurobi.com/}.
Table \ref{tab:simul} summarizes the main parameters used for simulation.

The algorithms that we consider in this section for evaluation are as follows.
\begin{itemize}
\item
\textbf{GreedyMRC:} 
The centralized MRC-based greedy algorithm proposed in \cite{MY04} introduced  in Section \ref{sec:StateArt}. 
Despite being centralized, due to lack of a more relevant work,  we use it as our main benchmark.

\item
\textbf{MC-MRC, SV-MRC:} 
Our proposed MRC-based algorithm with the MC and  the SV  schemes, respectively.

\item
\textbf{MC-OPN, SV-OPN:}
Special case of MC-MRC and SV-MRC in which the nodes can choose only one PN.
\item
\textbf{MILP-MRC, MILP-OPN:}
The MILP-based optimum solution obtained by solving \eqref{eq:milp_mrc_all} with and without exploiting MRC, respectively.

\end{itemize}

\subsection{Simulation results}

\begin{figure*}[!tp]
\centering
\subfloat[$p^{\mathrm{c}}= $ 1 mW]
{
\resizebox{.3  \textwidth}{!}
	{
   		\setlength\fheight{.55\columnwidth}  	
    	\setlength\fwidth{.65\columnwidth} 		
%
%
\definecolor{mycolor1}{rgb}{1.00000,0.00000,1.00000}%
\begin{tikzpicture}

\begin{axis}[%
width=0.951\fwidth,
height=\fheight,
at={(0\fwidth,0\fheight)},
scale only axis,
xmin=10,
xmax=25,
xlabel style={font=\color{white!15!black}},
xlabel={No. $N$ of Nodes},
ymin=2,
ymax=3.4,
ylabel style={font=\color{white!15!black}},
ylabel={Normalized network power  $\overbar{P}^\mathrm{tot}_\mathrm{net} (\bm{a})$},
axis background/.style={fill=white},
xmajorgrids,
ymajorgrids,
legend style={at={(0.03,0.97)}, anchor=north west, legend cell align=left, align=left, draw=white!15!black}
]
\addplot [color=mycolor1, dashed, line width=2.0pt, mark size=3.5pt, mark=square, mark options={solid, mycolor1}]
  table[row sep=crcr]{%
10	2.07420449220415\\
15	2.71260426475635\\
20	3.12593213112075\\
25	3.31418809129737\\
};
\addlegendentry{{\small GreedyMRC - \cite{MY04}}}

\addplot [color=red, line width=2.0pt, mark size=5.0pt, mark=o, mark options={solid, red}]
  table[row sep=crcr]{%
10	2.00822672185521\\
15	2.68615936885812\\
20	3.06389548033639\\
25	3.14143505629003\\
};
\addlegendentry{{\small MC-MRC}}

\end{axis}
\end{tikzpicture}%
  	}
   	\label{fig:netpow_low_pc} 
} 
\subfloat[$p^{\mathrm{c}}= $ 10 mW]
{
\resizebox{.3  \textwidth}{!}
	{
   		\setlength\fheight{.55\columnwidth}  	
    	\setlength\fwidth{.65\columnwidth} 		
%
%
\definecolor{mycolor1}{rgb}{1.00000,0.00000,1.00000}%
\begin{tikzpicture}

\begin{axis}[%
width=0.951\fwidth,
height=\fheight,
at={(0\fwidth,0\fheight)},
scale only axis,
xmin=10,
xmax=25,
xlabel style={font=\color{white!15!black}},
xlabel={No. $N$ of Nodes},
ymin=2,
ymax=9,
ylabel style={font=\color{white!15!black}},
ylabel={Normalized network power $\overbar{P}^\mathrm{tot}_\mathrm{net} (\bm{a})$},
axis background/.style={fill=white},
xmajorgrids,
ymajorgrids,
legend style={at={(0.03,0.97)}, anchor=north west, legend cell align=left, align=left, draw=white!15!black}
]
\addplot [color=mycolor1, dashed, line width=2.0pt, mark size=3.5pt, mark=square, mark options={solid, mycolor1}]
  table[row sep=crcr]{%
10	2.94349020648986\\
15	4.79188997904206\\
20	6.65688451207314\\
25	8.53847380558308\\
};
\addlegendentry{{\small GreedyMRC - \cite{MY04}}}

\addplot [color=red, line width=2.0pt, mark size=5.0pt, mark=o, mark options={solid, red}]
  table[row sep=crcr]{%
10	2.3490523309627\\
15	3.21510596668519\\
20	3.7850787243343\\
25	4.05897060391003\\
};
\addlegendentry{{\small MC-MRC}}

\end{axis}
\end{tikzpicture}%
  	}
   	\label{fig:netpow_med_pc} 
} 
\subfloat[$p^{\mathrm{c}}= $ 100 mW]
{
\resizebox{.3  \textwidth}{!}
	{
		\setlength\fheight{.55\columnwidth}  	
    	\setlength\fwidth{.65\columnwidth} 		
%
%
\definecolor{mycolor1}{rgb}{1.00000,0.00000,1.00000}%
\begin{tikzpicture}

\begin{axis}[%
width=0.951\fwidth,
height=\fheight,
at={(0\fwidth,0\fheight)},
scale only axis,
xmin=10,
xmax=25,
xlabel style={font=\color{white!15!black}},
xlabel={No. $N$ of Nodes},
ymin=0,
ymax=70,
ylabel style={font=\color{white!15!black}},
ylabel={Normalized network power  $\overbar{P}^\mathrm{tot}_\mathrm{net} (\bm{a})$},
axis background/.style={fill=white},
xmajorgrids,
ymajorgrids,
legend style={at={(0.03,0.97)}, anchor=north west, legend cell align=left, align=left, draw=white!15!black}
]
\addplot [color=mycolor1, dashed, line width=2.0pt, mark size=3.5pt, mark=square, mark options={solid, mycolor1}]
  table[row sep=crcr]{%
10	11.636347349347\\
15	25.5847471218992\\
20	41.9664083215969\\
25	60.7813309484402\\
};
\addlegendentry{{\small GreedyMRC - \cite{MY04}}}

\addplot [color=red, line width=2.0pt, mark size=5.0pt, mark=o, mark options={solid, red}]
  table[row sep=crcr]{%
10	5.42141078272783\\
15	8.00638648654349\\
20	10.1886492627552\\
25	11.9681991113629\\
};
\addlegendentry{{\small MC-MRC}}

\end{axis}
\end{tikzpicture}%
	}
   	\label{fig:netpow_high_pc} 
}  
\caption{Network power for different numbers $N$ of nodes in the network and different values $p^\mathrm{c}$ of  circuitry power.}
\label{fig:net_pow_mrc}
\end{figure*}
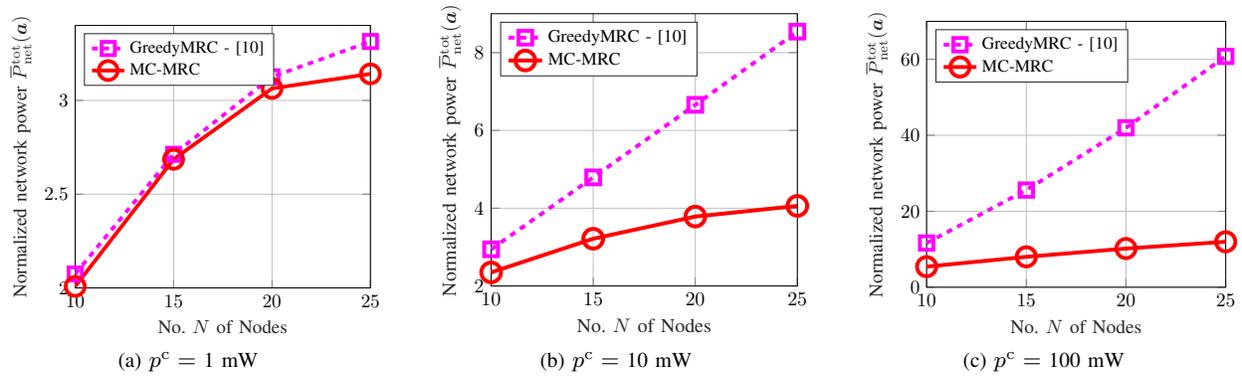

We first show the importance of taking the circuitry power into account for message dissemination.
We compare our proposed algorithm with GreedyMRC proposed in \cite{MY04} which is centralized, ignores the circuitry power required at the nodes and merely considers the power required for the radio link in network optimization.
Figures \ref{fig:netpow_low_pc}, \ref{fig:netpow_med_pc} and  \ref{fig:netpow_high_pc}, correspond to low, medium and high circuitry powers, respectively.
First, we observe that by increasing the number $N$ of nodes, the   network power increases.
This is because the circuitry powers required at the nodes for message reception and transmission impose additional power on the network which is not negligible.
Second, we observe in Fig. \ref{fig:netpow_low_pc} that by increasing the number of nodes, the powers required by both algorithms  tend to saturate.
When the network becomes denser, the number of PNs required for covering the network and serving all the receiving nodes does not necessarily increase.
Hence, by increasing the number of nodes, at some point, the total radio link power required in the network for message dissemination remains unchanged.
However, the circuitry power required at the nodes makes the network power continue to increase.
The value of the circuitry power in Fig. \ref{fig:netpow_high_pc} is higher than that for the other two cases.
Thus, in Fig. \ref{fig:netpow_high_pc}, even if the total radio-link power of the network does not change significantly, the high value of the circuitry power dominates the radio-link power required in the network.
This results in a constant increase of the network power in Fig. \ref{fig:netpow_high_pc}.

Our proposed algorithm outperforms GreedyMRC in all the three cases shown in Fig. \ref{fig:net_pow_mrc}.
When the circuitry power of the nodes is high, the performance of our game-theoretic algorithm becomes significantly better than the benchmark algorithm, see Fig. \ref{fig:netpow_high_pc}.
The main reason is that with our approach the nodes take the circuitry power into account in choosing their PNs. 
In contrast, with \cite{MY04}, the message is always transmitted over large number of hops in order to minimize the total radio link power of the network.
Since each transmission requires circuitry power besides the radio-link power, the actual network power required by \cite{MY04} for message dissemination, after adding the circuitry power of the nodes to the outcome of the algorithm, becomes significantly high.
This is more pronounced with high values of the circuitry power.

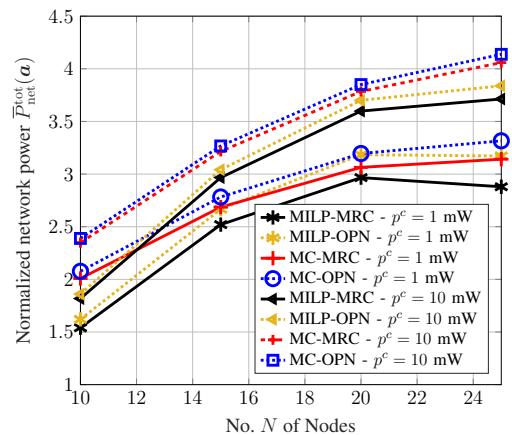
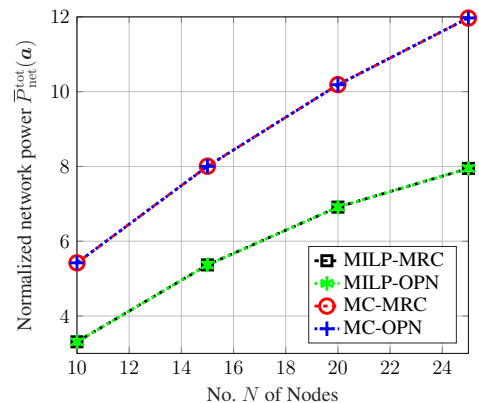
\begin{figure}[!t]
\centering
\subfloat[The network power with $p^\mathrm{c} \in \{1 \text{ mW}, 10 \text{ mW}\}$.]
{
\resizebox{.8  \columnwidth}{!}{
		\setlength\fheight{.50\textwidth}  	
    	\setlength\fwidth{.60\textwidth} 		
    	\begin{Large}
%
%
\definecolor{mycolor1}{rgb}{0.90000,0.70000,0.10000}%
\begin{tikzpicture}

\begin{axis}[%
width=0.951\fwidth,
height=\fheight,
at={(0\fwidth,0\fheight)},
scale only axis,
xmin=10,
xmax=25,
xlabel style={font=\color{white!15!black}},
xlabel={No. $N$ of Nodes},
ymin=1,
ymax=4.5,
ylabel style={font=\color{white!15!black}},
ylabel={Normalized network power  $\overbar{P}^\mathrm{tot}_\mathrm{net} (\bm{a})$},
axis background/.style={fill=white},
xmajorgrids,
ymajorgrids,
legend style={at={(0.97,0.03)}, anchor=south east, legend cell align=left, align=left, draw=white!15!black}
]
\addplot [color=black, line width=2.0pt, mark size=5.0pt, mark=asterisk, mark options={solid, black}]
  table[row sep=crcr]{%
10	1.54039275954441\\
15	2.52012300114611\\
20	2.96646200410263\\
25	2.87940976841399\\
};
\addlegendentry{{\large MILP-MRC - $p^c = 1$ mW}}

\addplot [color=mycolor1, dotted, line width=2.0pt, mark size=5.0pt, mark=asterisk, mark options={solid, mycolor1}]
  table[row sep=crcr]{%
10	1.62031492617235\\
15	2.66647898505248\\
20	3.18386073918723\\
25	3.17246018857661\\
};
\addlegendentry{{\large MILP-OPN - $p^c = 1$ mW}}

\addplot [color=red,  line width=2.0pt, mark size=5.0pt, mark=+, mark options={solid, red}]
  table[row sep=crcr]{%
10	2.00822672185521\\
15	2.68615936885812\\
20	3.06389548033639\\
25	3.14143505629003\\
};
\addlegendentry{{\large MC-MRC - $p^c = 1$ mW}}

\addplot [color=blue, dotted, line width=2.0pt, mark size=5.0pt, mark=o, mark options={solid, blue}]
  table[row sep=crcr]{%
10	2.07655656527955\\
15	2.78234896569646\\
20	3.19546159284319\\
25	3.31589444671972\\
};
\addlegendentry{{\large MC-OPN - $p^c = 1$ mW}}

\addplot [color=black, line width=2.0pt, mark size=3.3pt, mark=triangle, mark options={solid, rotate=90, black}]
  table[row sep=crcr]{%
10	1.81774608051566\\
15	2.96667085599008\\
20	3.59849446187833\\
25	3.71321689818041\\
};
\addlegendentry{{\large MILP-MRC - $p^c = 10$ mW}}

\addplot [color=mycolor1, dotted, line width=2.0pt, mark size=3.3pt, mark=triangle, mark options={solid, rotate=90, mycolor1}]
  table[row sep=crcr]{%
10	1.86020967365351\\
15	3.04136920517309\\
20	3.7007711452887\\
25	3.83841549400032\\
};
\addlegendentry{{\large MILP-OPN - $p^c = 10$ mW}}

\addplot [color=red, dashed, line width=2.0pt, mark size=3.5pt, mark=+, mark options={solid, red}]
  table[row sep=crcr]{%
10	2.3490523309627\\
15	3.21510596668519\\
20	3.7850787243343\\
25	4.05897060391003\\
};
\addlegendentry{{\large MC-MRC - $p^c = 10$ mW}}

\addplot [color=blue, dotted, line width=2.0pt, mark size=3.5pt, mark=square, mark options={solid, blue}]
  table[row sep=crcr]{%
10	2.38876875068498\\
15	3.2672750441538\\
20	3.85017438516119\\
25	4.13746677370717\\
};
\addlegendentry{{\large MC-OPN - $p^c = 10$ mW}}

	\end{axis}
\end{tikzpicture}%
    	\end{Large}
	}
   	\label{fig:pow_low_pc} 
}  
\\
\subfloat[The network power with $p^\mathrm{c} = 100$ mW.]{
\resizebox{.75  \columnwidth}{!}{
		\setlength\fheight{.45\textwidth}  	
    	\setlength\fwidth{.55\textwidth} 		
    	\begin{Large}
%
%
\begin{tikzpicture}

\begin{axis}[%
width=0.951\fwidth,
height=\fheight,
at={(0\fwidth,0\fheight)},
scale only axis,
xmin=10,
xmax=25,
xlabel style={font=\color{white!15!black}},
xlabel={No. $N$ of Nodes },
ymin=3,
ymax=12,
ylabel style={font=\color{white!15!black}},
ylabel={Normalized network power  $\overbar{P}^\mathrm{tot}_\mathrm{net} (\bm{a})$},
axis background/.style={fill=white},
xmajorgrids,
ymajorgrids,
legend style={at={(0.97,0.03)}, anchor=south east, legend cell align=left, align=left, draw=white!15!black}
]
\addplot [color=black, dashed, line width=2.0pt, mark size=3.5pt, mark=square, mark options={solid, black}]
  table[row sep=crcr]{%
10	3.309380967193\\
15	5.36617617252812\\
20	6.91238945987361\\
25	7.94802082922947\\
};
\addlegendentry{{\Large MILP-MRC }}

\addplot [color=green, dotted, line width=2.0pt, mark size=5.0pt, mark=asterisk, mark options={solid, green}]
  table[row sep=crcr]{%
10	3.30938096719331\\
15	5.36617617252891\\
20	6.91238945987446\\
25	7.94802082922996\\
};
\addlegendentry{{\Large MILP-OPN }}

\addplot [color=red, dashed, line width=2.0pt, mark size=5.0pt, mark=o, mark options={solid, red}]
  table[row sep=crcr]{%
10	5.42141078272783\\
15	8.00638648654349\\
20	10.1886492627552\\
25	11.9681991113629\\
};
\addlegendentry{{\Large	 MC-MRC }}

\addplot [color=blue, dotted, line width=2.0pt, mark size=5.0pt, mark=+, mark options={solid, blue}]
  table[row sep=crcr]{%
10	5.42141078272783\\
15	8.00254106735859\\
20	10.1863224106658\\
25	11.9727548126496\\
};
\addlegendentry{{\Large MC-OPN }}

\end{axis}
\end{tikzpicture}%
    	\end{Large}
	}
   	\label{fig:pow_high_pc} 
} 
\caption{The normalized network power for different numbers $N$ of nodes in the network and different values $p^\mathrm{c}$ of  circuitry power.}
\label{fig:netpow_mrc}
\end{figure}

In Fig. \ref{fig:netpow_mrc}  we  evaluate  the effect of employing the MRC in message dissemination.
We depict the network power obtained via the MC and the MILP approaches  for different numbers of the nodes and different values of the circuitry power.
First, we observe that with MRC, the message can be disseminated with a lower network power compared to the case of one PN (OPN).
Second, we find in Fig. \ref{fig:pow_low_pc}  that, when the  circuitry power is low  ($p^\mathrm{c}$ = 1 mW), employing the MRC technique results in a higher gain compared to the case in which the circuitry power is high.
In fact, when the radio-link power dominates the circuitry power, it is beneficial for the nodes to combine the signals from multiple PNs.
Conversely in Fig. \ref{fig:pow_high_pc}, where the circuitry power is high, i.e., 100 mW, the nodes do not exploit the MRC as it requires reception circuitry power over multiple time-slots.
In other words, as shown in Fig. \ref{fig:pow_high_pc},  even if the nodes are allowed to choose multiple PNs, they only select one PN and hence, the performance of the MPN and OPN approaches are almost the same.

\begin{figure}[!t]
\centering
\subfloat[The social cost with $p^\mathrm{c} \in \{1 \text{ mW}, 10 \text{ mW}\}$.]
{
\resizebox{.75 \columnwidth}{!}{
		\setlength\fheight{.55\textwidth}  	
    	\setlength\fwidth{.65\textwidth} 		
    	\begin{huge}
%
%
\begin{tikzpicture}

\begin{axis}[%
width=0.951\fwidth,
height=\fheight,
at={(0\fwidth,0\fheight)},
scale only axis,
xmin=10,
xmax=25,
xlabel style={font=\color{white!15!black}},
xlabel={No. $N$ of Nodes },
ymin=2,
ymax=3.8,
ylabel style={font=\color{white!15!black}},
ylabel={Normalized social cost $\overbar{\mathrm{SC}}  (\bm{a})$},
axis background/.style={fill=white},
xmajorgrids,
ymajorgrids,
legend style={at={(0.97,0.03)}, anchor=south east, legend cell align=left, align=left, draw=white!15!black}
]
\addplot [color=blue, dotted, line width=2.0pt, mark size=3.5pt, mark=square, mark options={solid, blue}]
  table[row sep=crcr]{%
10	2.14438535898102\\
15	2.89391178044325\\
20	3.31576762351188\\
25	3.40995288818689\\
};
\addlegendentry{{\Large SV-OPN - $p^c = 1$ mW}}

\addplot [color=red, dashed, line width=2.0pt, mark size=5.0pt, mark=o, mark options={solid, red}]
  table[row sep=crcr]{%
10	2.09531475661892\\
15	2.80408871511988\\
20	3.20395725516388\\
25	3.29492037675095\\
};
\addlegendentry{{\Large SV-MRC - $p^c = 1$ mW}}

\addplot [color=blue, dotted, line width=2.0pt, mark size=3.3pt, mark=triangle, mark options={solid, rotate=90, blue}]
  table[row sep=crcr]{%
10	2.28048291995663\\
15	3.11002295939332\\
20	3.61023164929564\\
25	3.78110898966359\\
};
\addlegendentry{{\Large SV-OPN - $p^c = 10$ mW}}

\addplot [color=red, dashed, line width=2.0pt, mark size=5.0pt, mark=+, mark options={solid, red}]
  table[row sep=crcr]{%
10	2.23851169834476\\
15	3.03012014693317\\
20	3.51142774743394\\
25	3.68243449984706\\
};
\addlegendentry{{\Large SV-MRC - $p^c = 10$ mW}}

\end{axis}
\end{tikzpicture}%
    	\end{huge}
	}
   	\label{fig:socost_low_pc} 
}
 
\caption{The normalized social cost for different numbers $N$ of nodes and different values $p^\mathrm{c}$ of circuitry power.}
\label{fig:socost_mrc}
\end{figure}
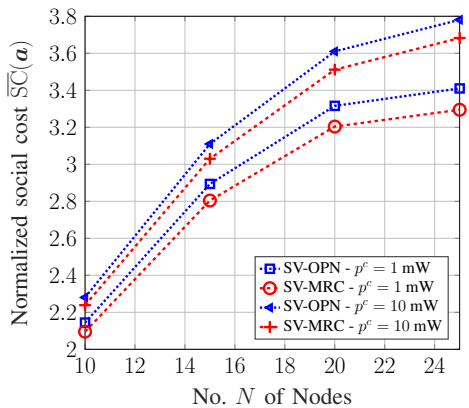

Figure \ref{fig:socost_mrc} evaluates the effect of employing MRC on the social cost.
As mentioned earlier, when the  social cost minimization is the objective,  the nodes do not consider their own reception circuitry power.
In fact, the goal of the nodes in this case is to minimize the power imposed on their chosen PNs and consequently to minimize the cost they pay.
Similar to the MC-based game in Fig. \ref{fig:netpow_mrc}, here, SV-MRC performs better that SV-OPN in terms of the social cost.
In other words,  the total cost paid by the receiving nodes in order to receive the message reduces  when they are allowed to receive the message from more than one PN.
By comparing Fig. \ref{fig:pow_low_pc} and Fig. \ref{fig:socost_low_pc}, we find that employing the MRC has a slightly higher gain on the SV-based game than the MC-based game.
The reason is that in the SV-based game,  the reception circuitry power  is not considered in the cost function of a CN. 
Hence, receiving via the MRC over multiple time-slots imposes a lower cost on the receiving nodes compared to the MC-based game in which the cost of reception is also included.
 
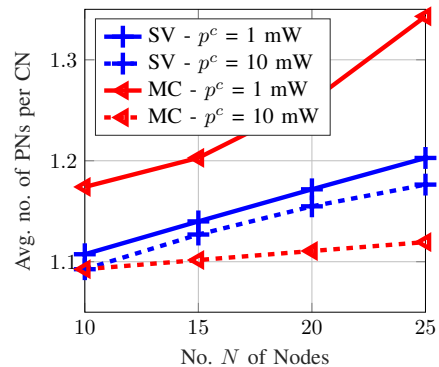
\begin{figure} [!t]
    \centering
    \resizebox{.7 \columnwidth}{!}{
    \setlength\fheight{.55\columnwidth}  	
    \setlength\fwidth{.65\columnwidth} 		
%
%
\begin{tikzpicture}

\begin{axis}[%
width=0.951\fwidth,
height=\fheight,
at={(0\fwidth,0\fheight)},
scale only axis,
xmin=10,
xmax=25,
xlabel style={font=\color{white!15!black}},
xlabel={No. $N$ of Nodes},
ymin=1.05,
ymax=1.35,
ylabel style={font=\color{white!15!black}},
ylabel={Avg. no. of PNs per CN},
axis background/.style={fill=white},
xmajorgrids,
ymajorgrids,
legend style={at={(0.03,0.97)}, anchor=north west, legend cell align=left, align=left, draw=white!15!black}
]
\addplot [color=blue, line width=2.0pt, mark size=5.0pt, mark=+, mark options={solid, blue}]
  table[row sep=crcr]{%
10	1.10740740740741\\
15	1.13993780748167\\
20	1.17172793093846\\
25	1.20277777777778\\
};
\addlegendentry{SV - $p^c$ = 1 mW}

\addplot [color=blue, dashed, line width=2.0pt, mark size=5.0pt, mark=+, mark options={solid, blue}]
  table[row sep=crcr]{%
10	1.09259259259259\\
15	1.12693771465701\\
20	1.15486981342245\\
25	1.17638888888889\\
};
\addlegendentry{SV - $p^c$ = 10 mW}

\addplot [color=red, line width=2.0pt, mark size=3.3pt, mark=triangle, mark options={solid, rotate=90, red}]
  table[row sep=crcr]{%
10	1.17407407407407\\
15	1.2028868467465\\
20	1.25921400724032\\
25	1.34305555555556\\
};
\addlegendentry{MC - $p^c$ = 1 mW}

\addplot [color=red, dashed, line width=2.0pt, mark size=3.3pt, mark=triangle, mark options={solid, rotate=90, red}]
  table[row sep=crcr]{%
10	1.09259259259259\\
15	1.1016662025434\\
20	1.11061681982735\\
25	1.11944444444444\\
};
\addlegendentry{MC - $p^c$ = 10 mW}

\end{axis}
\end{tikzpicture}%
    }
	\caption{The average number of PNs per CN for different numbers $N$ of nodes in the network.}
    \label{fig:mrc_pn_per_cn} 
\end{figure}

In order to find the effect of the MRC technique on nodes' decision,  in Fig. \ref{fig:mrc_pn_per_cn}  we show the average number of PNs per CN with two values for the circuitry power, 1 mW and 10 mW.
Considering the MC-MRC and the SV-MRC for low circuitry power, i.e., $p^\mathrm{c} = 1$ mW, we observe that the average number of PNs per CN for the MC-based game is higher than that for the SV-based game.
The reason is that, since the objective is  network power minimization and since the circuitry power in this case is low, the nodes exploit the MRC technique to a higher extent than the SV-based game in order to reduce their costs as well as the network power.

In addition, we observe that, in general, when the circuitry power increases, the average number of PNs per CN reduces for both the MC-based and the SV-based games.
This reduction is more significant for the MC-based game.
Recall that, with the MC-based game, the cost of a CN with respect to the \textit{ circuitry power} of its PN is either zero (in multicast) or $p^\mathrm{c}$ (in unicast), see \eqref{eq:mrc:mc_breakdown}.
Moreover, according to \eqref{eq:sv_with_pc}, with the SV-based game, the circuitry power of transmission at the PN is equally shared among the CNs.
In short, the impact of the transmission circuitry power on the  cost function of a node  is not as significant as the impact of the reception circuitry power which is only captured in the MC-based game.
Therefore, as the circuitry power increases from $p^\mathrm{c} = 1$ mW to $p^\mathrm{c} = 10$ mW, the nodes in the MC-based game react by choosing a lower number of PNs, while in contrast, such an increase of the circuitry power has a lower impact on the decision of the nodes in the SV-based game.
For instance, when there are 20 nodes in the network, by increasing the circuitry power from $p^\mathrm{c} = 1$ mW to $p^\mathrm{c} = 10$ mW, the average number of PN per CN with the MC-based game reduces from 1.26 to 1.11 PNs per CN.
These numbers for the SV-based game are 1.18 and 1.16 PNs per CN, respectively.

\begin{figure}[!t]
\centering
\includegraphics[width=.8  \columnwidth]{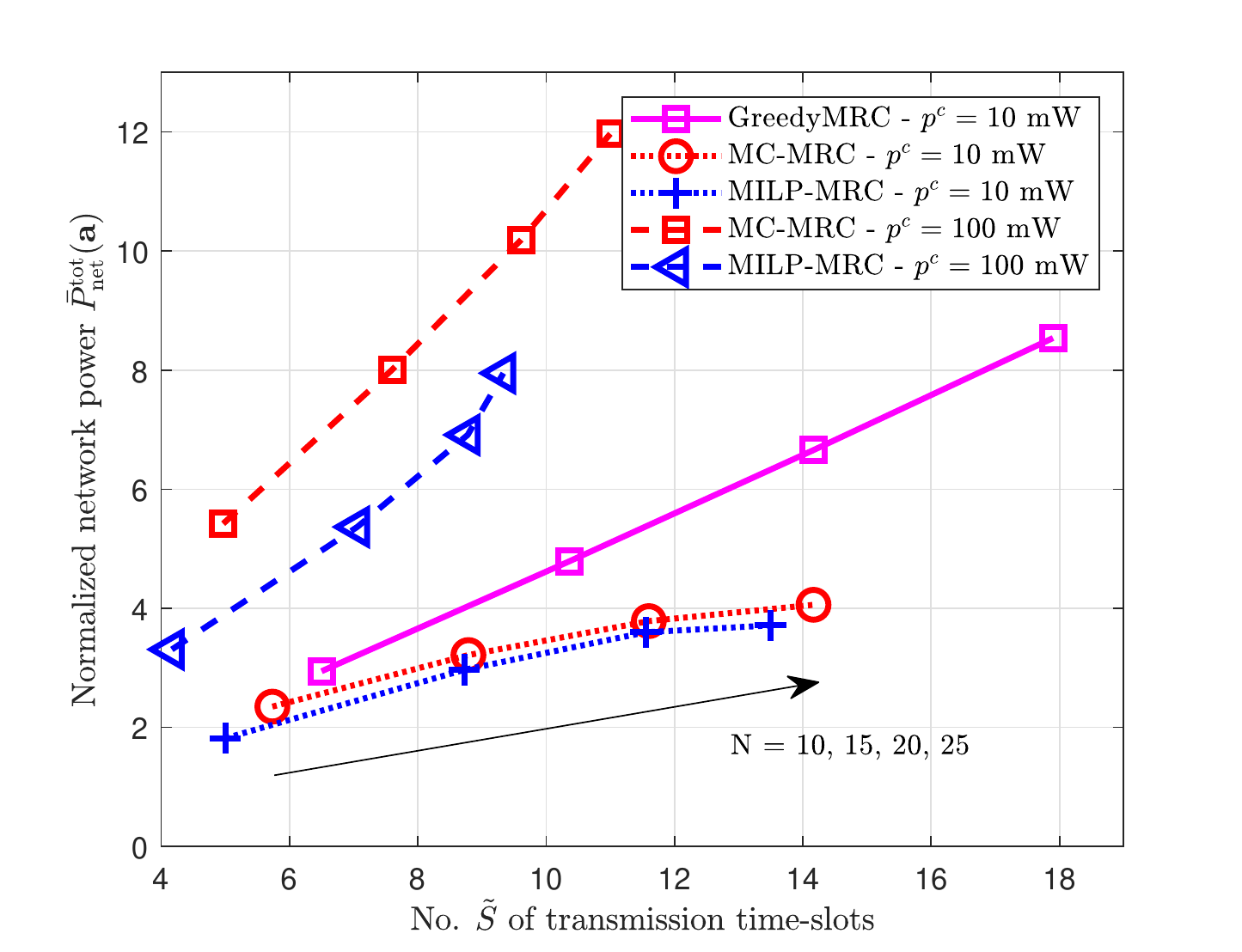}
\caption{Normalized network power $\overbar{P}^\mathrm{tot}_\mathrm{net} (\bm{a})$ versus the number $\tilde{S}$ of time-slots used for message dissemination. }
\label{fig:netpow_tslot} 
\end{figure}

Finally, Fig. \ref{fig:netpow_tslot} shows how efficient the resources are used in this network for different algorithms.
We plot the network power versus the number of time-slots used in the network for message dissemination.
The optimum MILP-based approach and the MC-based game are considered with two different values of circuitry power, that is, $p^c =$ 10 mW and $p^c =$ 100 mW.
Further, the performance of the GreedyMRC is also shown for the case of $p^c =$ 10 mW.
Since the network power obtained by the GreedyMRC for $p^c =$ 100 mW  is much higher than for the other algorithms, we omit the GreedyMRC for $p^c =$ 100 mW.
In Fig. \ref{fig:netpow_tslot}, the closer the points are to the origin, the better the resources are used.
As can be seen, 
when the circuitry power increases, the number of time-slots required for message dissemination in the network decreases.
In other words, transmissions  over a large number of hops impose additional transmission and reception circuitry powers on the nodes.
Hence, cost minimization at the nodes results in a message dissemination strategy that requires  a  lower number of time-slots.
Considering both Fig. \ref{fig:netpow_tslot} and Fig. \ref{fig:mrc_pn_per_cn}, it can be inferred that by increasing the circuitry power, the nodes tend to receive the message from one PN and  the multicast receiving groups are formed by a larger number of CNs so that both the average number of PNs per CN as well as the average number of required time-slots reduce.

\section{Summary}
\label{sec:summ_mPN}

We studied multi-hop broadcast in a network with one source and multiple receivers.
The nodes are able to exploit the MRC technique and combine the messages received from multiple transmitting nodes in order to reduce the network power required for message dissemination.
We studied two scenarios depending on the need for an incentive for the transmitting nodes.
A  decentralized approach using a non-cooperative CSG is proposed in which the nodes employ the MC and the SV cost sharing schemes for minimization of network power in incentive-independent networks and social cost in incentive-mandatory networks, respectively.
We showed that our proposed game is a potential game that possesses an NE.
Simulation results showed that our proposed algorithm  outperforms the existing heuristic algorithm concerning the required power  for message dissemination while it is also able to address the incentive and fairness issues.
Further, since our algorithm captures the impact of the circuitry power on the network power, the higher the circuitry power of the nodes, the higher gain is obtained by our algorithm compared to the benchmarks. 
We also presented the centralized global optimum via an MILP that can be used for both scenarios.

\begin{appendices}
\section{Proof of Lemma \ref{lem:sv_piecewise}}
\label{app:a}
Assume that $p_{i,j}^\mathrm{req}$ is the $(n+1)$th lowest request from PN $j$ as shown in \eqref{eq:pow_sort} such that  $p_{i,j}^\mathrm{req} = p_{n+1,j}^\mathrm{req}$.
Based on \eqref{eq:sv_radio} by considering $i=n+1$, $c_{j,i}^\mathrm{SV} (p_{i,j}^\mathrm{req},  \bm{p}_{-i,j}^\mathrm{rcv})$ can be written as a function of $n$ as
\begin{align}  
 \label{eq:cost_expand}
c^\mathrm{SV}_{j,i}  (p_{i,j}^\mathrm{req},\bm{p}_{-i,j}^\mathrm{rcv}) = & \frac{p_{i,j}^{\mathrm{req}} - p_{n,j}^{\mathrm{req}}}{ M_j +1-(n+1)}  + \sum_{k=1}^n \frac{p_{k,j}^{\mathrm{req}} - p_{k-1,j}^{\mathrm{req}}}{ M_j +1-k}.
\end{align}
Expanding the right side of \eqref{eq:cost_expand} leads to 
\begin{align} 
 \label{eq:cost_expand2}
c^\mathrm{SV}_{j,i} & (p_{i,j}^\mathrm{req},\bm{p}_{-i,j}^\mathrm{rcv}) =  \frac{ p_{i,j}^{\mathrm{req}} }{M_j-n} - \frac{ p_{n,j}^{\mathrm{req}}}{M_j-n} \notag \\
 & +  \frac{   p_{n,j}^{\mathrm{req}}}{M_j-n +1}   + \dots  - \frac{ p_{1,j}^{\mathrm{req}}}{M_j -1 } +  \frac{ p_{1,j}^{\mathrm{req}}}{M_j}.
\end{align}
Eq. \eqref{eq:cost_expand2} is equivalent to
\begin{align}
 \label{eq:SV2}
c^\mathrm{SV}_{j,i} &(p_{i,j}^\mathrm{req},\bm{p}_{-i,j}^\mathrm{rcv}) =  \frac{p_{i,j}^\mathrm{req}}{M_j-n} 
  + \sum\limits_{k=1}^{n \geq 1}  \left( \frac{ - p_{k,j}^\mathrm{req} } { (M_j-k ) (M_j-k+1)} \right).
\end{align}
It can be derived from \eqref{eq:SV2} that the cost of node $i$ is obtained by a linear function whose slope and y-intercept depend on the interval that $p_{i,j}^\mathrm{req}$ falls in.
Eq. \eqref{eq:SV2} shows that if $p^\mathrm{req}_{i,j}$  increases and falls inside an interval with a higher $n$, the slope of the function $c_{j,i} $  in \eqref{eq:SV2} increases accordingly.
Besides, the y-intercept of $c_{j,i} $   decreases.
Hence, $c_{j,i} $  forms a   piece-wise linear function that increases in the interval $[0, p_j^\mathrm{max}]$.
\end{appendices}

\bibliographystyle{IEEEtran}
\bibliography{ref}

\end{document}